\documentclass[11pt,letterpaper]{article}

\usepackage[margin=1in]{geometry}
\usepackage[T1]{fontenc}
\usepackage{lmodern}
\usepackage{microtype}
\usepackage{amsmath,amssymb,amsthm}
\usepackage{graphicx}
\usepackage{float}
\usepackage{array}
\usepackage{subcaption}   
\usepackage{algorithm}
\usepackage{algpseudocode}
\usepackage[title]{appendix}
\usepackage{enumitem}
\usepackage{cite}
\usepackage{xurl}
\usepackage[hidelinks]{hyperref}
\usepackage{cleveref}
\usepackage{fancyhdr}
\usepackage{orcidlink}

\numberwithin{equation}{section}

\newcolumntype{P}[1]{>{\centering\arraybackslash}p{#1}}
\newcommand{\funding}[1]{\par\medskip\noindent\textbf{Funding.} #1\par}
\newcommand{\roles}[1]{\par\medskip\noindent\textbf{Author contributions.} #1\par}
\newcommand{\data}[1]{\par\medskip\noindent\textbf{Data availability.} #1\par}

\let\ManuscriptIncludeGraphics\includegraphics
\renewcommand{\includegraphics}[2][]{%
  \IfFileExists{#2}{\ManuscriptIncludeGraphics[#1]{#2}}{%
    \fbox{\parbox[c][2.4cm][c]{0.67\linewidth}{%
      \centering\footnotesize Figure PDF not uploaded\\[-0.1em]
      \path{#2}}}%
  }%
}

\newcommand{\shorttitle}{Design Optimization of Nuclear Fusion Reactor}
\newcommand{\shortauthors}{Kim and Seo}
\renewcommand{\headrulewidth}{0.4pt}
\renewcommand{\footrulewidth}{0pt}
\fancypagestyle{plain}{%
  \fancyhf{}%
  \fancyfoot[C]{\thepage}%
  \renewcommand{\headrulewidth}{0pt}%
}

\begin{document}

\begin{center}
  {\LARGE\bfseries Design Optimization of Nuclear Fusion Reactor\\[0.1em]
   through Deep Reinforcement Learning\par}
  \vspace{1.25em}
  {\large Jinsu Kim$^{1}$\,\orcidlink{0009-0000-2610-4551} and
        Jaemin Seo$^{2,*}$\,\orcidlink{0000-0003-0635-0282}\par}
  \vspace{0.65em}
  {\small\itshape $^{1}$Department of Mechanical and Aerospace Engineering, Princeton University, United States\par}
  {\small\itshape $^{2}$Department of Physics, Chung-Ang University, Republic of Korea\par}
\end{center}
\begingroup
\renewcommand{\thefootnote}{\fnsymbol{footnote}}
\footnotetext[1]{Corresponding author: Jaemin Seo (\texttt{jseo@cau.ac.kr}).\quad Jinsu Kim: \texttt{jk9075@princeton.edu}.}
\endgroup
\vspace{1.1em}
\thispagestyle{plain}

\begin{abstract}
In this research, we investigate deep reinforcement learning (DRL) as a practical tool for constrained, multi-objective optimization for tokamak fusion reactor designs. The task is to optimize the design configurations that minimize building cost while achieving high energy gain and long energy confinement time, subject to the four operational limits required for stable operation. Due to the coupled physics-engineering evaluation, the resulting objectives and constraints are black-box, making it challenging to find optimal solutions with conventional search strategies. To address this, we implement Reward-Constrained Policy Optimization (RCPO) for tokamak fusion reactor design. This method integrates multiple constraints into the reward functions, guiding the policy towards feasible solutions and simplifying the design optimization process. Our numerical results show that the proposed DRL-based approach finds more feasible designs that satisfy operational constraints more efficiently than other data-driven optimization methods, demonstrating the potential of DRL for efficient future reactor design. 
\end{abstract}

\section{Introduction}
Nuclear fusion has been highlighted by demands for sustainability and clean energy, which have promoted efforts in building fusion power reactors globally. A magnetic confinement fusion reactor concept, called the Tokamak, is one of the promising designs that has demonstrated high performance across several aspects. However, optimizing fusion reactor design remains challenging due to operational constraints arising from instabilities and steady-state operation \cite{najmabadi1991aries, freidberg2015designing, kang2023massive}. These constraints are density limit \cite{greenwald1988new, greenwald2002density}, beta limit \cite{troyon1984mhd}, kink safety instability limit \cite{wesson1978hydromagnetic}, and the achievable bootstrap fraction \cite{kikuchi1990steady}. Plus, the reactor design should not exceed the maximum allowable neutron wall load and stress induced by electromagnetic forces on TF coils \cite{najmabadi1991aries, sborchia2008design, freidberg2015designing}. Given those constraints and conditions, the objective of fusion reactor design optimization includes reducing reactor costs \cite{jo2021cost, coleman2019blueprint}, increasing energy gain \cite{jo2021cost, jo2022design}, and improving confinement time. Thus, this task can be formulated as a multi-objective, multi-constraint optimization problem. 

The strategy for designing optimal fusion reactors has been investigated \cite{hong2018study, hong2019optimal, maisonnier2005conceptual, maisonnier2006demo, okano2000compact, najmabadi1991aries, jo2021cost, jo2022design, kang2023massive, freidberg2015designing, coleman2019blueprint, someya2018fusion}, but efficiently optimizing multi-objective constrained reactor designs remains a challenge. The difficulty of evaluating the design and computing the remaining design parameters, considering the wall load and stress, makes the problem a black-box objective function and constraints. Brute-force or random search algorithms have been utilized to find the optimal design that satisfies the constraints. However, this is inefficient for high-dimensional problems, and it is difficult to find feasible solutions under multiple constraints. A data-driven approach has been developed to address this problem, enabling the reduction of the high complexity associated with the underlying physical constraints. A Genetic Algorithm (GA) \cite{mirjalili2019genetic}, an evolutionary algorithm that solves optimization problems by generating candidate solutions to optimize by biologically inspired evolution operators such as selection, crossover, and mutation, has been used to optimize nuclear fission reactor design \cite{pereira2008parallel, kumar2015new, norouzi2013nuclear, sobolev2017genetic}. A Particle Swarm Optimization (PSO) \cite{kennedy1995particle}, a stochastic population-based optimization by iteratively improving a population (swarm) of candidate solutions (particles), which move through the search space using a combination of each particle's best-known position and the swarm's best-known position, has been applied to divertor design \cite{frerichs2025particle} and coil design \cite{wu2019design, zhao2021design, robles2018designing}. Bayesian Optimization (BO) \cite{garnett2023bayesian}, a Bayesian probabilistic approach to optimize the objective functions by building a surrogate model for the objective using Gaussian process regression and optimizing an acquisition function to get the best sample, has been employed to optimize the magnetic field coils \cite{nunn2023shaping, nunn2025bayesian} and superconducting magnets \cite{packman2025bayesian}. Recently, Deep Reinforcement Learning (DRL), which combines artificial neural networks with Reinforcement Learning (RL) \cite{sutton1999reinforcement}, has demonstrated remarkable generalization to unseen data. This provides the optimal policy that maximizes the cumulative reward associated with the system's objective through exploitation and exploration, making it well-suited for not only the active control tasks \cite{seo2021feedforward, seo2022development, degrave2022magnetic, seo2024avoiding} but also repeating optimization tasks, including design optimization \cite{yonekura2019framework}. Several studies have shown the validity of DRL in airfoil design optimization \cite{yonekura2019framework,rabault2020deep,viquerat2021direct,dussauge2023reinforcement} and strain design optimization \cite{sabzevari2022strain}.

In this research, we investigate the use of DRL to optimize tokamak design and verify that the DRL framework can efficiently discover optimal designs. Our goal is to find the design configurations that (1) minimize building cost, (2) achieve high energy gain, and (3) maximize energy confinement time, while simultaneously satisfying four operational limits required for stable operation. DRL provides a simplified search procedure by encoding design objectives and constraints into a scalarized reward, enabling direct exploration of the design space without high computational costs. To enforce feasibility during optimization, we adopt Reward-Constrained Policy Optimization (RCPO) \cite{tessler2018reward}, a constrained policy optimization method that integrates constraints as penalty terms into the reward function, thereby guiding the policy towards feasible solutions. Our approach begins with Proximal Policy Optimization (PPO) \cite{schulman2017proximal} as a baseline and integrates it with our reactor design computation code. By combining PPO with the RCPO technique, we enable efficient exploration of the feasible solution space to identify optimal designs. Our numerical results demonstrate that this approach finds optimal designs that satisfy operational constraints more efficiently than other data-driven methodologies. 

This paper is organized as follows. The tokamak design computation process and the design objectives to be optimized will be introduced. We then present our proposed optimization method and its adaptation to reactor design optimization. Finally, we will present the numerical results and compare the design performance to validate our approach. 

\section{Fusion Reactor Design Model}
To design fusion tokamak reactors, it is necessary to consider plasma physics, engineering, and nuclear physics \cite{freidberg2015designing, someya2018fusion, kang2025physics}. Following the system-level design procedure proposed by Freidberg et al. \cite{freidberg2015designing}, we developed a fusion reactor design computation code that determines the geometry, the plasma parameters, and the performance of the designed steady-state tokamak reactor from a small set of design variables. 

\begin{figure}[H]
\centerline{\includegraphics[width=0.8\textwidth, scale = 1.0]{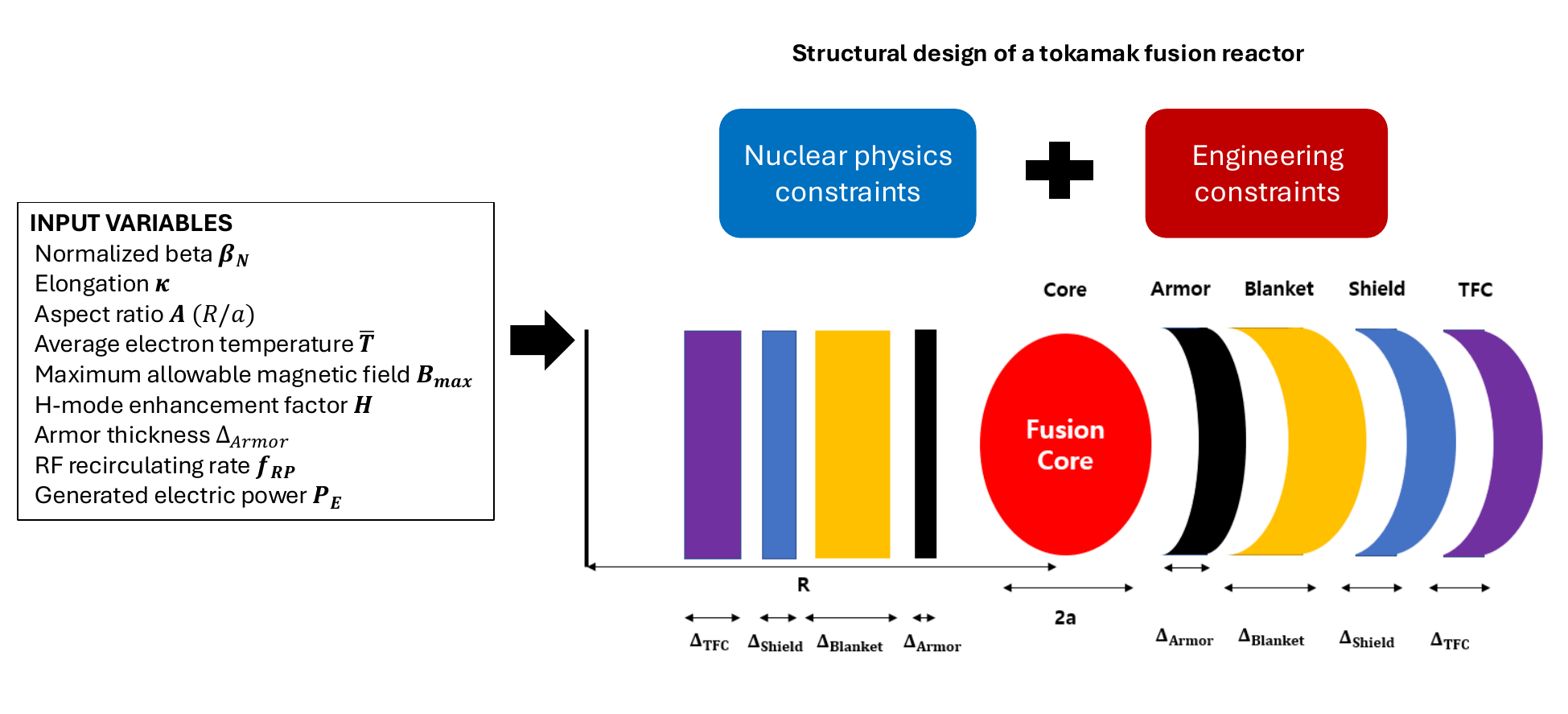}}
\caption{A simplified process of a tokamak fusion reactor design computation}
\label{fig:design-computation-scheme}
\end{figure}

The computation consists of four steps, as illustrated in Figure \ref{fig:design-computation-scheme}: (1) the geometrical design, (2) the plasma parameters, (3) the current drive and bootstrap current for steady-state operation, and (4) the evaluation and the operational limits. Note that the code considers the maximum allowable current density and stress on materials \cite{najmabadi1991aries, sborchia2008design, freidberg2015designing} in addition to neutron radiation wall loading \cite{freidberg2015designing, someya2018fusion} to determine the designs concerning the physically achievable construction of a reactor. We describe the details for each step in the following subsections. 

\subsection{Design variables}
The nine design variables listed in Table \ref{table:list-input-parameters} are the quantities required for calculating the geometrical design of the reactor, plasma parameters, and the designed reactor's performance. These are the input design variables to execute the computation and will be used as actions for the RL-based optimizer. 

\begin{table}[H]
\captionof{table}{A list of input design variables required for reactor design computation}
\begin{center}
\begin{tabular}{p{3cm}|c}
\hline
\multicolumn{1}{c|}{\textbf{Quantity}} & \textbf{Description}      \\ \hline
\multicolumn{1}{c|}{$\beta_\text{N}$} & Normalized ratio of plasma thermal and magnetic pressure\\
\multicolumn{1}{c|}{$\kappa$} & Elongation \\
\multicolumn{1}{c|}{$A$} & Aspect ratio \\
\multicolumn{1}{c|}{$\bar{T}$} & Average electron temperature \\
\multicolumn{1}{c|}{$B_{\max}$} & Maximum allowable magnetic field \\
\multicolumn{1}{c|}{$H$} & Confinement enhancement factor \\
\multicolumn{1}{c|}{$\Delta_{\text{Armour}}$} & Armour thickness \\
\multicolumn{1}{c|}{$f_{\text{RP}}$} & RF recirculating rate \\
\multicolumn{1}{c|}{$P_\text{E}$} & Generated electric power \\
\hline
\end{tabular}
\end{center}
\label{table:list-input-parameters}
\end{table}

If the input design variables are given, the computation code provides the output design variables listed in Table \ref{table:list-output-parameters}. All other parameters, including the maximum neutron wall load, the thermal conversion efficiency, the material limits of the TF coil, and the nuclear properties of the blanket, are held constant for every candidate design following \cite{freidberg2015designing}.

\begin{table}[H]
\captionof{table}{A list of output design variables determined by reactor design computation}
\begin{center}
\begin{tabular}{p{3cm}|c}
\hline
\multicolumn{1}{c|}{\textbf{Quantity}} & \textbf{Description}      \\ \hline
\multicolumn{1}{c|}{$R$} & Major radius\\
\multicolumn{1}{c|}{$a$} & Minor radius\\
\multicolumn{1}{c|}{$\Delta_{\text{Blanket}}$} & Blanket thickness\\
\multicolumn{1}{c|}{$\Delta_{\text{TFcoil}}$} & TF coil thickness\\
\multicolumn{1}{c|}{$\tau_\text{E}$} & Energy confinement time\\
\multicolumn{1}{c|}{$\beta$} & Plasma beta\\
\multicolumn{1}{c|}{$I_\text{p}$} & Plasma current\\
\multicolumn{1}{c|}{$q$} & Safety factor\\
\multicolumn{1}{c|}{$f_{\text{op}}$} & Required bootstrap fraction\\
\multicolumn{1}{c|}{$f_{\text{bs}}$} & Achievable neoclassical bootstrap fraction\\
\multicolumn{1}{c|}{$Q_{\parallel}$} & Parallel heat flux conducted to divertor\\
\multicolumn{1}{c|}{$\bar{n}$} & Average plasma density\\
\multicolumn{1}{c|}{$\bar{p}$} & Average plasma pressure\\
\multicolumn{1}{c|}{$n_\text{G}$} & Greenwald density\\
\multicolumn{1}{c|}{$\beta_{\text{Troyon}}$} & Troyon beta\\
\multicolumn{1}{c|}{$Q$} & Gaining factor\\
\multicolumn{1}{c|}{$C$} & Cost parameter\\
\hline
\end{tabular}
\end{center}
\label{table:list-output-parameters}
\end{table}

\subsection{Geometrical design}
Starting from the core design, the input variables from Table \ref{table:list-input-parameters} are required. The geometrical design is computed for the radial build of the reactor, including the major and minor radii, the blanket thickness, and the TF coil thickness.

\paragraph{Major and minor radius} The major radius of the core plasma can be determined through the neutron wall loading constraint. In more detail, the neutron wall loading at the first wall should not exceed the maximum allowable neutron wall load $P_W$. From the electric power $P_E$ and thermal efficiency $\eta_T$, the approximate neutron power at the wall can be estimated as $P_n=(E_n / E_F)P_E/\eta_T$, where $E_n$ is the neutron energy and $E_F$ is the total energy released per fusion reaction, including the energy from the breeding reaction with Li-6. The plasma surface area is approximated as $S \simeq 4\pi^2 R a [(1+\kappa^2)/2]^{0.5}$. Thus, by setting $P_n = P_\text{W} S$, the major radius can be computed as in Equation \ref{eq:major-radius}.

\begin{equation}
R = [\frac{1}{4\pi^2}\frac{E_n}{E_F}\frac{P_E}{\eta_T P_W}(\frac{2}{1 + \kappa ^ 2})^{0.5} A]^{0.5}
\label{eq:major-radius}
\end{equation}

Equation \eqref{eq:major-radius} assumes a maximum neutron wall load of $P_W \simeq 4 \text{MW}/\text{m}^2$ \cite{freidberg2015designing} with $E_n = 14.1 \text{MeV}$ and $E_F = 22.4 \text{MeV}$. In addition, the geometrical variables are also utilized, including elongation $\kappa$ and aspect ratio $A$, to compute the surface area and the minor radius from $A = R/a$.

\paragraph{Blanket thickness} The computation of blanket thickness is based on a simple planar structure with the governing equations for neutron slowing down and lithium breeding. Slowing down is explained as the collision of neutrons with Li-7 in the blanket, while the breeding process is the reaction between neutrons and Li-6. Thus, the energy balance equation for the collisional process and the mass balance equation from the nuclear reaction can be described by Equation 2 in \cite{freidberg2015designing}, enabling the estimation of the blanket thickness given by Equation \eqref{eq:blanket}. From the hard-sphere collision model, the scattering cross-section is estimated to be $\sigma_s \simeq 2 \text{barns}$, thus $\lambda_s = 0.1 \text{m}$ \cite{chadwick2011endf}. In addition, $\alpha_B = \lambda_B/\lambda_S(E_F / E_T)^{0.5}$ while $E_T = 0.025 \text{eV}$ \cite{chadwick2011endf}. $\Gamma_b/\Gamma_0$ is a ratio between the inward and outward fluxes of the blanket, assumed to be $10^{-5}$ \cite{freidberg2015designing}.

\begin{equation}
\Delta_{\text{Blanket}} = \lambda_s (1 + \alpha_B \ln {\frac{\Gamma_0}{\Gamma_b}})
\label{eq:blanket}
\end{equation}

\paragraph{Magnetic field.} The toroidal magnetic field decreases with the major radius as $1/R$, and the maximum allowable magnetic field $B_{\max}$ occurs at the inner leg of the TF coil, which is located at $R - a - \Delta_{\text{Blanket}}$. Thus, the magnetic field at the plasma center, denoted as $B_0$, is given by Equation \eqref{eq:magnetic-field} and is used to compute the plasma parameters in the following steps.

\begin{equation}
B_0 = B_{\max}(1 - \epsilon_\text{B}), \quad \epsilon_\text{B} = \frac{a + \Delta_{\text{Blanket}}}{R}
\label{eq:magnetic-field}
\end{equation}

\paragraph{TF coil thickness} We suppose the TF coil is a flat circular magnet for simplicity. The TF coil should maintain its strength against material stress from centering and tensile force from the magnetic field. After the simple calculation for deriving each component of the stresses \cite{freidberg2015designing}, the TF coil thickness is estimated as in Equation \eqref{eq:TF-coil}.

\begin{equation}
\Delta_{\text{TF}} = R \left\{2(1-\epsilon_B) - [(1-\epsilon_B)^2-\alpha_M]^{0.5} - [(1-\epsilon_B)^2-\alpha_J]^{0.5}\right\}
\label{eq:TF-coil}
\end{equation}

where $\alpha_M$ and $\alpha_J$ are defined as below. $\sigma_{max} = 600 \text{MPa}$ is the maximum allowable material stress applied to the coil, and $J_{max} = 20 \text{MA} /\text{m}^2$ is the maximum allowable overall current density for the superconducting coils \cite{najmabadi1991aries, sborchia2008design, freidberg2015designing}. 

\begin{equation}
\begin{split}
& \alpha_M = \frac{B_0^2}{\mu_0\sigma_{\text{max}}}[\frac{2\epsilon_B}{1+\epsilon_B} + \frac{1}{2}\ln(\frac{1+\epsilon_B}{1-\epsilon_B})]\\ 
& \alpha_J = \frac{2B_0}{\mu_0 R J_{\text{max}}}
\label{eq:alpha-M}
\end{split}
\end{equation}

\subsection{Plasma parameters}
After the geometric design computation, it is now possible to estimate the plasma parameters required to calculate the design performance. In this step, the average plasma temperature $\bar{T}$ and the enhancement factor $H$ are used as input variables to determine the remaining parameters. 

\paragraph{Plasma profiles.} Following \cite{freidberg2015designing}, we assume $T_e = T_i = T$, so that the plasma pressure is $p = 2nT$. The temperature, density, and pressure profiles are given as in Equation \eqref{eq:profiles}, where $\rho$ is the normalized minor radius and $\nu_p = \nu_T + \nu_n$. 

\begin{equation}
T = \bar{T}(1+\nu_T)(1-\rho^2)^{\nu_T}, \quad n = \bar{n}(1+\nu_n)(1-\rho^2)^{\nu_n}, \quad p = \bar{p}(1+\nu_p)(1-\rho^2)^{\nu_p}
\label{eq:profiles}
\end{equation}

\paragraph{Plasma pressure, density, and beta} From the profiles above and the fusion reactivity $\langle \sigma v \rangle$ computed from the fitting formula of the temperature as in  Equations 20 and 21 of \cite{freidberg2015designing}, the average plasma pressure is determined so that the thermal power generated by the fusion reactions produces the electric power $P_\text{E}$, and the related variables, including density and beta, are then determined sequentially. We can compute the average plasma pressure, density, and plasma beta, as given in Equations \eqref{eq:plasma-pressure}, \eqref{eq:plasma-density}, and \eqref{eq:plasma-beta}.

\begin{equation}
\bar{p} = [\frac{4}{\pi^2}(\frac{1+\nu_T}{1+\nu_P})^2 \frac{P_E}{\eta_T E_F R a^2 \kappa} \frac{\bar{T}^2}{\int_0^1 (1-\rho ^2)^{2\nu_n}\langle\sigma v \rangle \rho d\rho}]^{0.5}
\label{eq:plasma-pressure}
\end{equation}

\begin{equation}
\bar{n} = \frac{1+\nu_p}{2(1+\nu_T)(1+\nu_n)}\frac{\bar{p}}{\bar{T}}
\label{eq:plasma-density}
\end{equation}

\begin{equation}
\beta = \frac{2\mu_0\bar{p}}{B_0^2}
\label{eq:plasma-beta}
\end{equation}

\paragraph{Energy confinement time} From the power balance relation and empirical energy-confinement-time estimates for ELMy H-mode, the alpha heating power $P_\alpha = (E_\alpha/E_\text{F})(P_\text{E}/\eta_\text{T})$ balances the thermal conduction loss $3\pi^2 R a^2 \kappa \bar{p}/\tau_\text{E}$ in steady state. Thus, the energy confinement time is given by Equation \eqref{eq:confinement-time}.

\begin{equation}
\tau_E = 3 \pi ^ 2 R a^2 \kappa \frac{E_F \eta_T}{E_\alpha P_E}\bar{p}
\label{eq:confinement-time}
\end{equation}

\paragraph{Plasma current and safety factor.} By matching the energy confinement time in Equation \eqref{eq:confinement-time} with the empirical energy confinement time estimated for ELMy H-mode, we can derive the relation for the plasma current with the energy confinement time. First, the empirical energy confinement time is given by the IPB98(y,2) scaling law in Equation \eqref{eq:ipb98}.

\begin{equation}
\tau_\text{E} = 0.145 H I_\text{p}^{0.93} R^{1.39} a^{0.58} \kappa^{0.78} \bar{n}^{0.41} B_0^{0.15} M^{0.19} P_\alpha^{-0.69}
\label{eq:ipb98}
\end{equation}

Here, $I_\text{p}$ is in MA, $\bar{n}$ is in $10^{20}$ m$^{-3}$, $P_\alpha$ is in MW, and $M = 2.5$ is the average ion mass number of D-T plasma. Then, the plasma current is determined as in Equation \eqref{eq:plasma-current}, and the safety factor for the kink instability is computed as in Equation \eqref{eq:safety-factor}.

\begin{equation}
I_\text{p} = \frac{7.98}{H^{1.08}} \frac{\tau_\text{E}^{1.08}}{R^{1.49} a^{0.62} \kappa ^ {0.84} \bar{n}^{0.44} B_0 ^ {0.16} M ^ {0.2}} \left(\frac{E_\alpha P_\text{E}}{E_\text{F} \eta_\text{T}}\right)^{0.74}
\label{eq:plasma-current}
\end{equation}

\begin{equation}
q = \frac{2\pi a^2 B_0}{\mu_0 R I_\text{p}}\left(\frac{1+\kappa^2}{2}\right)
\label{eq:safety-factor}
\end{equation}

\paragraph{Energy gain.} The energy gain $Q$ is computed from the zero-dimensional power balance with the average plasma density and temperature. Here, $S_\text{F}$, $S_\alpha$, $S_\kappa$, and $S_{\text{rad}}$ are the fusion power density, alpha heating power density, thermal conduction loss, and radiation loss from bremsstrahlung and cyclotron radiation \cite{wesson2011tokamaks}, respectively. Then, $Q$ is given as the ratio of the fusion power to the external heating power required to sustain the plasma, as shown in Equation \ref{eq:energy-gain}.

\begin{equation}
Q = \frac{S_\text{F}}{S_\kappa + S_{\text{rad}} - S_\alpha}
\label{eq:energy-gain}
\end{equation}

\subsection{Current drive and bootstrap current}
For steady-state operation, the plasma current should be sustained by the external current drive and the neoclassical bootstrap current without the inductive current. Thus, the required bootstrap fraction for steady-state operation and the neoclassical bootstrap fraction \cite{wesson2011tokamaks} are estimated through Equations \eqref{eq:required-bs} and \eqref{eq:neoclassical-bs}, which are also suggested in \cite{freidberg2015designing}.

\paragraph{Required bootstrap fraction} The current drive is assumed to be provided by the lower hybrid wave. The required bootstrap fraction $f_{\text{op}}$ is the fraction of the plasma current that cannot be driven by the external current drive, as in Equation \eqref{eq:required-bs}.

\begin{equation}
f_{\text{op}} = 1 - 1.2\frac{P_{\text{CD}}}{R \bar{n} I_p n_{\parallel}^2}
\label{eq:required-bs}
\end{equation}

Here, $P_{\text{CD}}$ is the required current drive power, as $P_{CD} = \eta_{\text{RF}} f_{\text{RP}} P_E$, where $\eta_{RF}$ and $f_{\text{RP}}$ correspond to RF efficiency and RF recirculating rate, respectively. $n_{\parallel}$ is a parallel index of refraction, described as Equation \eqref{eq:parallel-index}.

\begin{equation}
n_{\parallel} \simeq \frac{\omega_{pe}}{\Omega_{e}} + \left[1 + \left(\frac{\omega_{pe}}{\Omega_{e}}\right)^2\right]^{0.5}\left[1 - \left(\frac{\omega_{\text{LH}}}{\omega}\right)^2\right]^{0.5}
\label{eq:parallel-index}
\end{equation}

$\omega_{pe}$ and $\Omega_{e}$ are a plasma oscillation frequency and a cyclotron frequency of electrons, while $\omega_{\text{LH}} = [\omega_{pi}^2/(1 + (\omega_{pe}/\Omega_{e})^2)]^{0.5}$ and $\omega_{pi}$ is a plasma oscillation frequency of ions. Following \cite{freidberg2015designing}, the wave frequency is set to $\omega = 2\omega_{\text{LH}}$, and these frequencies are evaluated at $\rho = 0.8$, where the wave is assumed to be absorbed.

\paragraph{Neoclassical bootstrap fraction} The achievable bootstrap fraction is estimated from the neoclassical bootstrap current density in a large aspect ratio tokamak \cite{wesson2011tokamaks}, given by Equation \eqref{eq:local-bs}.

\begin{equation}
J_\text{B} = -2.44\left(\frac{r}{R}\right)^{0.5}\frac{p}{B_\theta}\left(\frac{1}{n}\frac{\partial n}{\partial r} + \frac{0.055}{T}\frac{\partial T}{\partial r}\right)
\label{eq:local-bs}
\end{equation}

The elongated cross-section is replaced by a circular cross-section with the same area, whose radius is $\kappa^{0.5}a$. To compute the poloidal magnetic field $B_\theta$, the total current density profile is assumed as in Equation \eqref{eq:current-profile}, which has an off-axis peak due to the lower hybrid current drive \cite{freidberg2015designing}.

\begin{equation}
J(\rho) = \frac{I_\text{p}}{\pi \kappa a^2}\frac{9}{8}\rho^{0.25}\frac{\alpha^2(1-x)e^{\alpha x}}{e^\alpha - 1 - \alpha}
\label{eq:current-profile}
\end{equation}

By integrating $J_\text{B}$ over the cross-section with the given profiles, we can compute the neoclassical bootstrap fraction as in Equation \ref{eq:neoclassical-bs}.

\begin{equation}
f_{\text{bs}} = 268 \frac{a^{2.5}\kappa^{1.25}\bar{p}}{\mu_0 R^{0.5}I_p^2}\int_0^1{\frac{\rho^{2.5}(1-\rho^2)^{0.5}}{b_\theta}}d\rho
\label{eq:neoclassical-bs}
\end{equation}

Here, $b_\theta(\rho) = \frac{1}{\rho}[\frac{(1+\alpha-\alpha x)e^{\alpha x} - 1 - \alpha}{e^\alpha- 1 - \alpha}]$ is the normalized poloidal magnetic field obtained from Equation \eqref{eq:current-profile}, where $x = \rho ^ {\frac{9}{4}}$ and $\alpha = 2.53$.

\subsection{Evaluation and operational limits}
All basic parameters have been provided so far, enabling computation of design performance, including cost parameters, heat flux, and operational limits. 

\paragraph{Cost parameter} Generally, a larger tokamak reactor incurs higher costs but can generate higher power. Thus, the volume-to-generated-power ratio can be used to estimate cost \cite{spears1980scaling}. The cost parameter is then defined as in Equation \eqref{eq:cost}, where $V_B$ is a blanket volume and $V_{TF}$ is a TF coil volume.

\begin{equation}
C = \frac{V_B + V_{TF}}{P_E}
\label{eq:cost}
\end{equation}
Here, the volumes are calculated by Equation \eqref{eq:volumes}.
\begin{equation}
\begin{split}
& V_\text{B} = 2\pi^2 R\left[(a+\Delta_{\text{Blanket}})(\kappa a+\Delta_{\text{Blanket}}) - \kappa a^2\right]\\
& V_{\text{TF}} = 4\pi \Delta_{\text{TFcoil}} (2R - 2a - 2\Delta_{\text{Blanket}} - \Delta_{\text{TFcoil}})\left[(1+\kappa)a + 2\Delta_{\text{Blanket}} + \Delta_{\text{TFcoil}}\right]
\end{split}
\label{eq:volumes}
\end{equation}

\paragraph{Parallel heat flux} The heat flux component parallel to the magnetic field, which can be dissipated by contact with the divertor, should be considered. Equation \eqref{eq:heat-flux} depicts the estimated amount of parallel heat flux. In this work, $Q_{\parallel}$ is computed for each design but is not used as a constraint.

\begin{equation}
Q_{\parallel} = \frac{E_\alpha P_E}{E_F \eta_T} \frac{B_0}{R}
\label{eq:heat-flux}
\end{equation}

\paragraph{Operational limits} Finally, the density limit and beta limit can be computed as Equations \eqref{eq:density-limit} and \eqref{eq:beta-limit}, respectively.

\begin{equation}
\bar{n}_{G} = \frac{I_p}{\pi a^2}
\label{eq:density-limit}
\end{equation}

\begin{equation}
\beta_{\text{Troyon}} = \beta_N \frac{I_p}{aB_0}
\label{eq:beta-limit}
\end{equation}

The following evaluation process is then required to verify the feasibility of the reactor design. In the sense of plasma physics, the four operational constraints \cite{freidberg2015designing} to avoid instabilities and to achieve steady-state operation must be satisfied: density limit \cite{greenwald1988new, greenwald2002density}, beta limit \cite{troyon1984mhd}, kink stability factor limit \cite{wesson1978hydromagnetic}, and bootstrap fraction limit \cite{kikuchi1990steady}. Density limit \cite{greenwald1988new, greenwald2002density} and Beta limit \cite{troyon1984mhd} refer to the safety margin of the plasma density and beta (i.e., the ratio of plasma pressure to magnetic pressure) for stable operation, respectively. The kink safety factor limit is the condition for avoiding the current-driven major MHD instability. The bootstrap fraction limit states that the achievable neoclassical bootstrap fraction $f_{\text{bs}}$ should exceed the required bootstrap fraction $f_{\text{op}}$ for steady operation. 

\begin{enumerate}
    \item Density limit: $\bar{n} < n_{G} = \frac{I_p}{\pi a^2}$
    \item Beta limit:$\beta < \beta_{\text{Troyon}} = \beta_\text{N} \frac{I_p}{aB_0}$
    \item Kink safety factor limit: $q > q_{\text{kink}} = 2$
    \item Bootstrap fraction limit: $f_{\text{bs}} > f_{\text{op}}$
\end{enumerate}

To achieve practical energy generation, three objectives are considered: minimizing building cost $C$, achieving high energy gain $Q$, and maximizing energy confinement time $\tau_\text{E}$.

\begin{enumerate}
    \item Low building cost $C$
    \item High energy confinement time ($\tau_\text{E}$)
    \item High $Q$ for efficiency
\end{enumerate}

Several studies have been conducted \cite{hong2018study, hong2019optimal, kim2015conceptual, jo2021cost, jo2022design, kang2023massive} to determine the optimal reactor design that minimizes cost while satisfying operational constraints \cite{jo2021cost}. We aim to optimize a tokamak reactor design to meet the minimum requirements for steady-state operation \cite{freidberg2015designing} while minimizing cost \cite{jo2021cost, jo2022design} through deep reinforcement learning.

\section{Design Optimization through Deep Reinforcement Learning}
Design optimization requires exploring a high-dimensional design parameter space, which entails high computational cost for high-fidelity numerical simulation \cite{viquerat2021direct, dussauge2023reinforcement}. Two main approaches have been used for optimizing design parameters: gradient-based and gradient-free methods. Gradient methods use the gradient of the design objective functions, and the adjoint method can reduce the gradient computation \cite{papadimitriou2008aerodynamic}, but they are easily trapped in local optima and are sensitive to the initial points \cite{viquerat2021direct}. On the other hand, gradient-free methods, including genetic algorithms \cite{samad2008shape}, are efficient at finding global optima and less sensitive to the choice of initial point, yet have a higher computational cost than gradient-based methods \cite{skinner2018state, viquerat2021direct}. Machine-learning-based surrogate model methods \cite{mack2007surrogate} have been highlighted to address the computational burden of simulations and improve design performance. Still, they depend on the dataset's quality \cite{dussauge2023reinforcement}.

Recent studies have demonstrated the effectiveness of deep reinforcement learning in design optimization fields \cite{yonekura2019framework, rabault2020deep, viquerat2021direct, dussauge2023reinforcement}. Reinforcement Learning (RL), a data-driven approach for learning to make decisions through trial and error \cite{sutton1999reinforcement}, provides a high generalization capability to unobserved system configurations \cite{yonekura2019framework}. Deep Reinforcement Learning (DRL) \cite{sewak2019deep}, combined with Deep Neural Networks (DNNs) and RL, has achieved high performance across diverse fields \cite{wang2022deep}. Since neural networks can approximate arbitrary nonlinear functions, guaranteed by the universal approximation theorem \cite{hornik1989multilayer}, integrating deep neural networks with reinforcement learning enables agents to efficiently learn an optimal policy while handling unstructured, high-dimensional input data. Here, we used Proximal Policy Optimization (PPO) \cite{schulman2017proximal} combined with a Reward-Constrained Policy Optimization (RCPO) for a constrained multi-objective design optimization problem in Tokamak fusion reactors. PPO is a policy-gradient method \cite{sutton1999policy} to optimize reactor design, which has several benefits, including the ability to handle continuous action spaces and stability through robust optimization by incorporating a form of trust-region optimization \cite{schulman2015trust}. RCPO is a multi-timescale constrained policy optimization \cite{tessler2018reward}, incorporating the penalty term for a constraint into the reward function. This allows the policy to explore the action space in a more constraint-aware way without reducing exploration efficiency.

\begin{figure}[H]
\centerline{\includegraphics[width=0.8\textwidth, scale = 1.0]{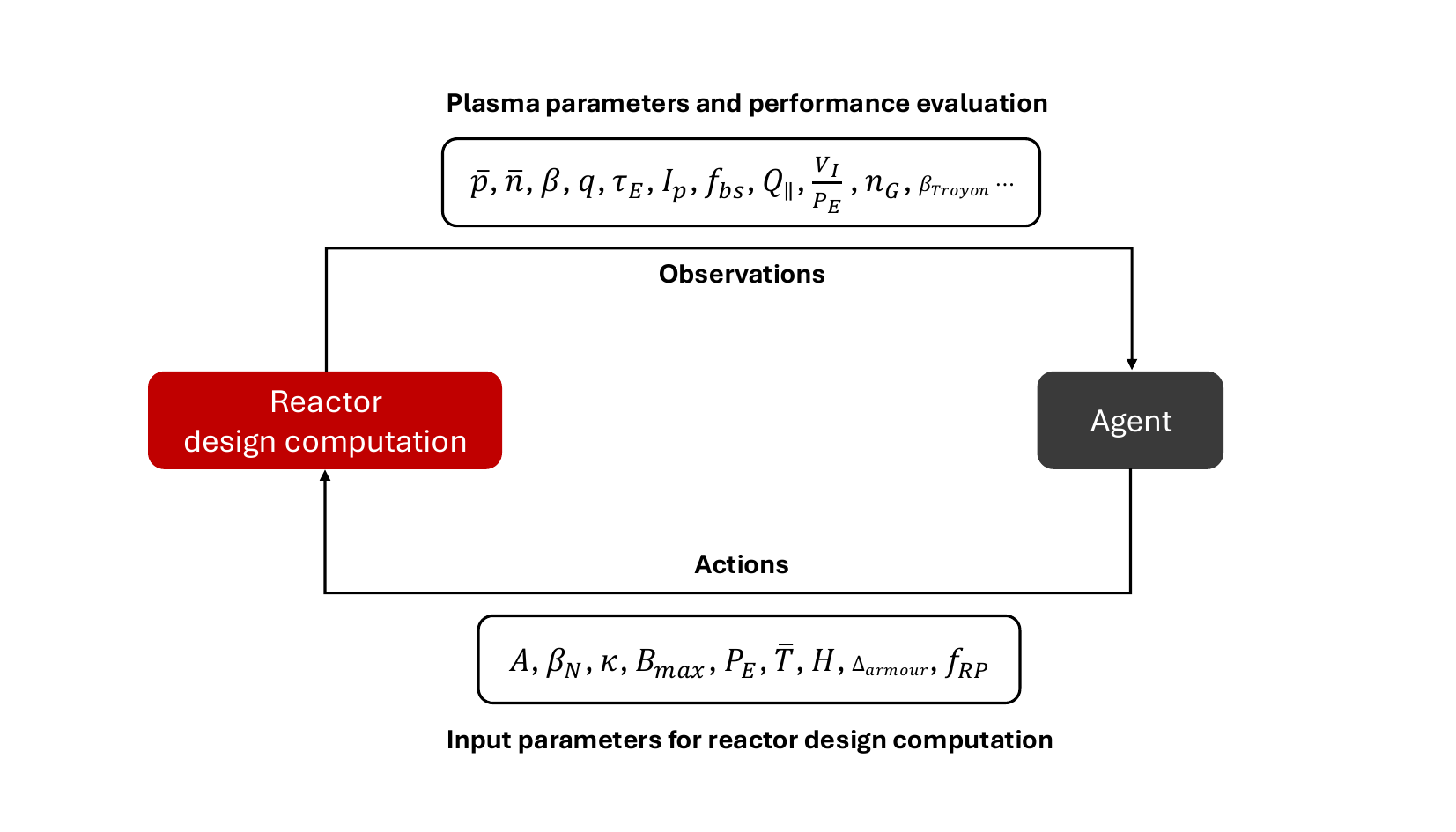}}
\caption{The proposed framework for reactor design optimization}
\label{fig:design-optimization-scheme}
\end{figure}

Figure \ref{fig:design-optimization-scheme} illustrates our proposed framework for Tokamak fusion reactor design optimization using deep reinforcement learning. We built the environment based on the reactor design computation code. Then, we trained the actor-critic network to find the optimal input parameters, as listed in Table \ref{table:list-input-parameters}. The environment provides the plasma parameters as the next state and evaluates the reward based on the design performance. The reactor design is then optimized during training, in which the agent learns optimal input parameters by being rewarded for its actions. More details of the proposed algorithm are presented in the following section.

\subsection{Proximal Policy Optimization (PPO)}
Our reactor design optimization builds on Proximal Policy Optimization (PPO) \cite{schulman2017proximal} with an Actor-Critic architecture. PPO is an improved version of Trust Region Policy Optimization (TRPO) \cite{schulman2015trust}: whereas TRPO maximizes a surrogate objective subject to a KL-divergence constraint on policy change, PPO replaces this with a clipped policy ratio, yielding the clipped surrogate objective

\begin{equation}
L^{\mathrm{CLIP}}(\theta) = \mathbb{E}_t\!\left[\min\!\left(r_t(\theta)\hat{A}_t,\;\mathrm{clip}(r_t(\theta), 1-\epsilon, 1+\epsilon)\hat{A}_t\right)\right],
\label{eq:clip-surrogate-objective}
\end{equation}

\noindent where $r_t(\theta) = \pi_\theta(a_t|s_t)/\pi_{\theta_{\mathrm{old}}}(a_t|s_t)$ is the probability ratio between the updated and old policies, $\hat{A}_t$ is an estimator of the advantage function at timestep $t$, and $\epsilon$ is a clipping hyperparameter. The advantage estimate is defined as
\begin{equation}
\hat{A}_t = -V(s_t) + r_t + \gamma r_{t+1} + \cdots + \gamma^{T-t-1} r_{T-1} + \gamma^{T-t} V(s_T).
\end{equation}

Since the Actor-Critic network shares parameters between the policy and value estimator, the full training objective combines the policy surrogate, a value function error term $L_t^{VF}(\theta) = \bigl(V_\theta(s_t) - V_t^{\mathrm{target}}\bigr)^2$, and an entropy bonus $S[\pi_\theta] = \mathbb{E}_{a\sim\pi_\theta}[-\log \pi_\theta(a_t|s_t)]$ that encourages exploration \cite{haarnoja2017reinforcement}:

\begin{equation}
L^{\mathrm{PPO}}(\theta) = \mathbb{E}_t\!\left[L_t^{\mathrm{CLIP}}(\theta) - c_1 L_t^{VF}(\theta) + c_2 S[\pi_\theta](s_t)\right],
\label{eq:ppo-loss}
\end{equation}

\noindent where $c_1$ and $c_2$ are hyperparameters. Algorithm~\ref{alg:ppo} summarizes the standard PPO procedure.

\begin{algorithm}[H]
\caption{Proximal Policy Optimization (PPO)}\label{alg:ppo}
\begin{algorithmic}
\For{episode $= 1, \ldots, N$}
    \State Run policy $\pi_\theta$ for $T$ timesteps and collect trajectory samples
    \State Estimate advantage values $\hat{A}_t$ using $V_\phi$ for $t = 1, \ldots, T$
    \State Optimize the surrogate loss $L^{\mathrm{PPO}}$ over the collected samples
    \State Update $\theta$ and $\phi$
\EndFor
\end{algorithmic}
\end{algorithm}

\subsection{Reward engineering for a reactor design optimization}
Designing a reward function is important to apply a deep reinforcement learning framework to the optimization task. Here, we have three different design objectives: (1) minimization of the building cost $C$, (2) maximization of the energy confinement time $\tau_E$, and energy gain $Q$. Plus, there are four different constraints that are the minimum requirements for steady-state operation. Thus, we will consider the reward function for each design objective and then express the constraints for steady-state operation with a Lagrange multiplier. 

For each design objective, we define reference parameters ($1/C_{\text{ref}}$, $\tau_{E,\text{ref}}$, $Q_{\text{ref}}$) corresponding to the parameters for the reference design proposed by \cite{freidberg2015designing}. For each parameter, we define a sigmoid function $\sigma(x):= \frac{\exp(x)}{\exp(x)+1}$ as a function of the ratio for the parameters ($1/C$,$\tau_E$, $Q$). This enables the designed reactor to have better performance than a reference design.

\begin{equation}
R(x) := \sigma({\frac{x}{x_{\text{ref}}}})
\label{eq:reward-function}
\end{equation}

Next, we express the constraints for the steady-state operation as the ratio of the plasma parameters to their maximum allowable values, such as Greenwald density $n_G$ and Troyon beta $\beta_\text{Troyon}$, as well as their minimum required values, including $q_{\text{kink}}$ and $f_{\text{op}}$, which are denoted by $(C_G, C_\beta, C_{q}, C_{\text{bs}})$. 

\begin{align}
\text{Density limit:} \quad & C_G = \frac{\bar{n}}{n_{\text{G}}} - 1 \\
\text{Beta limit:} \quad & C_{\beta} = \frac{\beta}{\beta_{\text{Troyon}}} - 1 \\
\text{Kink safety factor limit:} \quad & C_{q} = \frac{q_{\text{kink}}}{q} - 1 \\
\text{Bootstrap fraction limit:} \quad & C_{\text{bs}} = \frac{f_{\text{op}}}{f_{\text{bs}}} - 1
\end{align} These constraints determine the feasibility of the solutions found by the optimizers. Note that a design is feasible only if all four constraint residuals are negative. 

One of the easiest ways to evaluate the solutions for multi-objective optimization problems is the weighted sum of the evaluations for each objective. In this work, we define the weighted sum of each reward function for ($1/C,\tau_\text{E},Q$) with the feasibility terms corresponding to steady-state operation conditions $(C_G, C_\beta, C_q,C_{\text{bs}})$ as \ref{eq:total-reward}. 

\begin{equation}
R_{\text{total}} = w_{c} R_{\text{C}} + w_{\tau_\text{E}} R_{\tau_\text{E}} + w_{Q} R_{Q} -\lambda_{G}C_G-\lambda_\beta C_\beta - \lambda_{q}C_q - \lambda_{\text{bs}}C_{\text{bs}}
\label{eq:total-reward}
\end{equation} where $(w_C,w_{\tau_\text{E}},w_Q)$ are the weight values and $(\lambda_G, \lambda_\beta, \lambda_q,\lambda_{\text{bs}})$ are the Lagrange multipliers for the constraints. By doing so, the feasibility of the solutions is also considered, resulting in the optimal reactor design that satisfies steady-state operation conditions and maximizes the total reward. 

\subsection{Reward-Constrained Policy Optimization (RCPO)}
Standard PPO maximizes expected cumulative reward without explicit constraint satisfaction guarantees. To enforce operational constraints in reactor design, we incorporate Reward-Constrained Policy Optimization (RCPO) \cite{tessler2018reward}, a framework for solving constrained optimization problems. RCPO considers a constrained Markov decision process (CMDP), in which the policy maximizes expected cumulative reward while satisfying the constraints. Let $J_\pi^R=\mathbb{E}^\pi_{s\sim \mu}[\sum_{t=0}^{\infty}\gamma^t r_t]$ denote the expectation of the discounted cumulative reward, and let $J_\pi^C=\mathbb{E}^\pi_{s\sim \mu}[\sum_{t=0}^{\infty}C(s_t)]$ be the expected cumulative constraint cost, where $C(s_t)$ is a constraint. Then, the constrained policy optimization problem is written as
\begin{equation}
\max_{\pi}{J_\pi^R} \quad \text{subject to} \quad J_\pi^C \leq \alpha
\label{eq:constrained-policy-optimization}
\end{equation} 
where $\alpha \in [0,1]$ is a threshold for a constraint $C(s_t)$. 

The main idea of RCPO is to convert the constrained problem into an unconstrained one using a Lagrangian relaxation. For a parameterized policy $\pi_{\theta}$, the resulting objective is
\begin{equation}
\min_{\lambda} \max_{\theta}{[J_{\pi_\theta}^R - \lambda \cdot (J_{\pi_\theta}^C - \alpha)]}
\label{eq:unconstrained-policy-optimization}
\end{equation}
where $\lambda$ is a set of nonnegative Lagrange multipliers. In this Lagrangian relaxation, policies that violate the constraint incur a larger penalty term in $J_{\pi}^C$ and become less favorable during training. Therefore, this method balances reward maximization and constraint satisfaction within a single optimization framework. As discussed in \cite{tessler2018reward}, the solution of the unconstrained policy optimization problem \eqref{eq:unconstrained-policy-optimization} approaches that of the original constrained policy optimization problem \eqref{eq:constrained-policy-optimization} under the Lagrangian formulation. 

A key feature of RCPO is that the penalty term is adaptive, with $\lambda$ as a learnable parameter. This leads to a coupled optimization procedure for $\theta$ and $\lambda$ with two timescales as in \ref{eq:gradient-parameter}. 
\begin{align}
\lambda & \leftarrow \Gamma_{\lambda} (\lambda - \eta_\lambda \nabla_{\lambda} L(\theta,\lambda)) = \Gamma_{\lambda} (\lambda + \eta_\lambda \mathbb{E}^{\pi_\theta}_{s\sim \mu}[C(s)-\alpha]) \\
\theta & \leftarrow \Gamma_{\theta} (\theta + \eta_\theta \nabla_{\theta} L(\theta,\lambda)) = \Gamma_{\theta} (\theta + \eta_\theta \mathbb{E}^{\pi_\theta}_{s\sim \mu}[\nabla_\theta \log \pi(s,a;\theta) (R(s)-\lambda \cdot C(s))])
\label{eq:gradient-parameter}
\end{align}
where $\Gamma_\theta$ and $\Gamma_\lambda$ are projection operators for $\theta$ and $\lambda$ respectively (See \cite{tessler2018reward} for the details). On the faster timescale, the policy parameters $\theta$ are updated by maximizing the penalized objective in \ref{eq:unconstrained-policy-optimization}, just as in a standard policy optimization algorithm. Then, the Lagrange multiplier $\lambda$ is updated on a slower timescale according to the amount of constraint violation. If the current policy violates the constraint excessively, $\lambda$ increases, strengthening the penalty in subsequent iterations; if the policy is feasible, the penalty term relaxes, allowing the algorithm to search for a saddle-point $(\theta^*,\lambda^*)$ of \ref{eq:unconstrained-policy-optimization}, which is a high-performing feasible solution \cite{tessler2018reward}.

\subsection{PPO-Based RCPO algorithm for Reactor Design Optimization}
We now describe how PPO and RCPO are combined together for fusion reactor design optimization (Algorithm~\ref{alg:design-optimization}). Because the reactor design problem is evaluated at a single operating point rather than over a trajectory, we adopt a single-step RL formulation: at each episode, the agent selects a set of input design parameters as its action (Table \ref{table:list-input-parameters}), the environment computes the resulting reactor design and plasma parameters as the next state, and a reward incorporating both design objectives and constraint costs is returned immediately.

The PPO clipped surrogate objective \ref{eq:ppo-loss} is used to update the policy, while the RCPO Lagrangian relaxation \ref{eq:unconstrained-policy-optimization} augments the reward with an adaptive penalty $\lambda \cdot C(s)$ for constraint violations. After collecting $T$ timesteps of samples in a trajectory buffer $\tau$, the policy network $\pi_\theta$ and value network $V_\phi$ are updated jointly using RMSProp, and $\lambda$ is updated according to the average constraint violation observed in the batch. Between updates, the trajectory buffer is reinitialized from the reference design parameters to stabilize training in the single-step setting.

\begin{algorithm}[H]
\caption{PPO-Based RCPO for Fusion Reactor Design Optimization}\label{alg:design-optimization}
\begin{algorithmic}
\State Initialize policy $\pi_\theta$, value function $V_\phi$, and Lagrange multiplier $\lambda \geq 0$
\State Initialize state and action from reference design parameters:
\State \quad $s_0 \gets (R, a, \Delta_{\mathrm{Blanket}}, \Delta_{\mathrm{TFcoil}}, \tau_
\mathrm{E}, \beta, I_p, q, f_{\mathrm{bs}}, f_{\mathrm{op}}, Q_\parallel, \bar{n}, \bar{p}, n_G, \beta_T, Q, C)_{\mathrm{ref}}$
\State \quad $a_0 \gets (\beta_\mathrm{N}, \kappa, A, \bar{T}, B_{\max}, H, \Delta_{\mathrm{Armour}}, \eta_{\mathrm{RF}}, P_\mathrm{E})_{\mathrm{ref}}$
\State Compute $r_0 \gets R(s_0, a_0)$; add $(s_0, a_0, r_0)$ to $\tau$
\State $r_{\max} \gets r_0$,\quad $t \gets 0$
\For{each episode $t = 0, 1, \ldots, N-1$}
    \State $a_{t+1} \sim \pi_\theta(s_t)$ \Comment{Take action from current policy}
    \State $s_{t+1} \gets \mathrm{Env}(a_{t+1})$ \Comment{Compute new reactor design and plasma parameters}
    \State $r_{t+1} \gets R(s_{t+1}, a_{t+1}) - \lambda \cdot C(s_{t+1})$ \Comment{Penalized reward from RCPO}
    \State Add $(s_{t+1}, a_{t+1}, r_{t+1})$ to $\tau$;\quad $t \gets t + 1$
    \If{$t = T$}
        \State Update $\pi_\theta$ and $V_\phi$ by optimizing $L^{\mathrm{PPO}}$ on $\tau$ \Comment{Policy update}
        \State $\lambda \leftarrow \Gamma_\lambda\!\left(\lambda + \eta_\lambda\,\frac{1}{T}\sum_{t'}\bigl[C(s_{t'}) - \alpha\bigr]\right)$ \Comment{Lagrange multiplier update}
        \State Reinitialize $s_t$, $a_t$, $r_t$ from reference design;\quad $t \gets 0$
    \EndIf
    \If{$r_{t} \geq r_{\max}$}
        \State Optimal design $\gets (s_t, a_t)$ \Comment{Save best feasible design}
        \State $r_{\max} \gets r_t$
    \EndIf
\EndFor
\end{algorithmic}
\end{algorithm}

Note that the proposed framework simplifies the constrained multi-objective design optimization problem by using a weighted sum of the rewards for each design objective, while the Lagrangian relaxation technique yields high-performing feasible solutions. In more detail, this approach does not require explicit derivatives of each design objective with respect to the input parameters, unlike gradient-based multi-objective optimization. This is particularly useful in the present reactor design problem, where the design variables, plasma quantities, and engineering constraints are strongly coupled through a black-box computational model, while reducing computational costs in multi-objective optimization and reflecting the user's design preferences by controlling the weights. In the next section, we will demonstrate the effectiveness of the proposed method for reactor design optimization.

\section{Numerical Results}
\subsection{Experimental setup}
In this experiment, we employed the Reward-Constrained Policy Optimization (RCPO) algorithm built on Proximal Policy Optimization (PPO) \cite{schulman2017proximal} as the proposed method, denoted as DRL for convenience. As baseline optimization methods, we considered Grid Search (GS), Genetic Algorithm (GA), Particle Swarm Optimization (PSO), and Bayesian Optimization (BO). For GA, the population size was set to $32$, the mutation rate was $0.4$, and the mutation standard deviation was $1.96$. In PSO, the number of particles was set to $32$, and the update weight was $0.3$. In BO, we used a Gaussian Process Regressor (GPR) implemented in scikit-learn with the L-BFGS optimizer, and the most recent $256$ samples were used to optimize the surrogate model at each iteration. All methods were run for 5,000 iterations, which was sufficient to observe saturation in the design objective evaluation during optimization. To evaluate the design, we set $w_\text{C}=w_{\tau_{\text{E}}}=w_\text{Q}=1/3$ and $\lambda=0.25$ for all cases. 

For the RL-based method, both the actor and critic networks consisted of three hidden layers with 64 dimensions in each layer. The standard deviation of the policy distribution used for action sampling was fixed at 0.5. As the loss function for PPO, we used the weighted sum of the policy surrogate loss, combined with the CLIP loss and entropy bonus term, while the value function was trained using SmoothL1Loss \cite{schulman2017proximal}. The clipping parameter for CLIP loss $\epsilon$ was $0.2$, and the entropy coefficient was $0.05$. The learning rate for the network's parameters $\theta$ was set to $2.0 \times 10^{-4}$, and the discount factor $\gamma$ was 0.999. The learning rate for the Lagrange multipliers $\lambda$ was $2.0 \times 10^{-4}$, and we set the initial value $\lambda=1.0$ for each constraint. The optimizer used in this research was RMSProp for both $\theta$ and $\lambda$.

\begin{table}
\begin{center}
\begin{tabular}{|p{3cm}|c}
\hline
\multicolumn{1}{c|}{\textbf{Input parameter}} & \textbf{Search space}      \\ \hline
\multicolumn{1}{c|}{$\beta_N$} & [2.0, 4.0]\\
\multicolumn{1}{c|}{$\kappa$} & [1.5, 1.9] \\
\multicolumn{1}{c|}{$A$} & [3.5, 4.5] \\
\multicolumn{1}{c|}{$\bar{T}$ (KeV)} & [10, 15] \\
\multicolumn{1}{c|}{$B_{\max}$ (T)}& [10, 16] \\
\multicolumn{1}{c|}{$H$}& [1.0, 1.3] \\
\multicolumn{1}{c|}{$\Delta_{\text{Armour}}$ (m)}& [0.01, 0.05] \\
\multicolumn{1}{c|}{$f_{\text{RP}}$} & [0.1, 0.2] \\
\multicolumn{1}{c|}{$P_\text{E}$ (MW)}& [500, 1500] \\
\hline
\end{tabular}
\end{center}
\captionof{table}{A list of the search space for each input parameter corresponding to Table \ref{table:list-input-parameters}.}
\label{table:input-parameters-range}
\end{table}

The search space for input parameters was bounded to ensure efficient and stable optimization. Table \ref{table:input-parameters-range} describes the range of each input parameter allowed during the optimization. For GS, the input parameters were randomly sampled from a Gaussian distribution over these ranges at each iteration. For GA, PSO, BO, and RL, the optimization process was divided into two stages: an offline stage ($n_{\text{iter}} < 1000$) and an online stage ($n_{\text{iter}} > 1000$). During the offline stage, candidate design configurations were randomly generated from a Gaussian distribution with a mean and standard deviation restricted to these ranges. In GA and PSO, these initial samples were used to form a feasible candidate pool for subsequent search. In BO and RL, these samples were used to initialize the Gaussian process regressor and the critic network, respectively, to more accurately estimate the evaluations of design candidates during the online stage. In particular, we utilized the reward-weighted regression method \cite{peters2007reinforcement, peng2019advantage} for offline RL training. 

\subsection{Optimization process}

\begin{figure}
     \centering
     \begin{subfigure}[b]{0.475\textwidth}
         \centering
         \includegraphics[width=\textwidth, scale = 1.0]{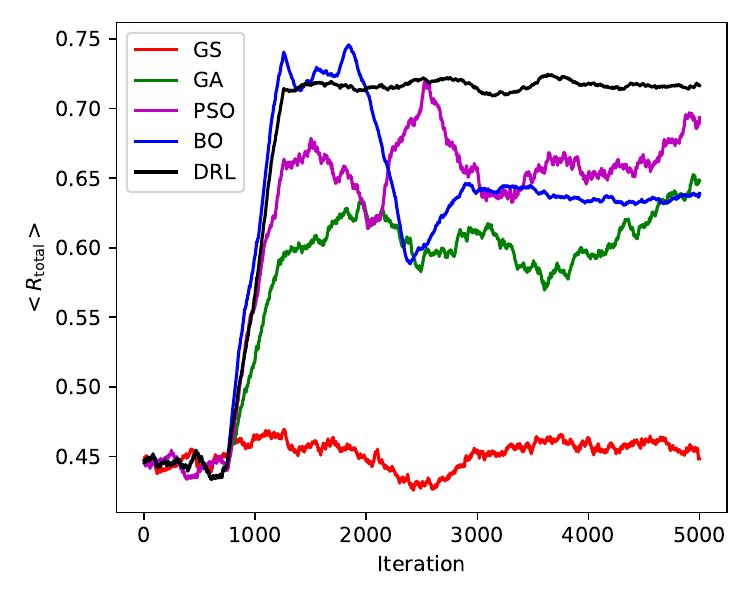}
         \caption{Total reward $R_{\text{total}}$}
         \label{fig:optimization-process}
     \end{subfigure}
     \hfill
     \begin{subfigure}[b]{0.475\textwidth}
         \centering
         \includegraphics[width=\textwidth, scale = 1.0]{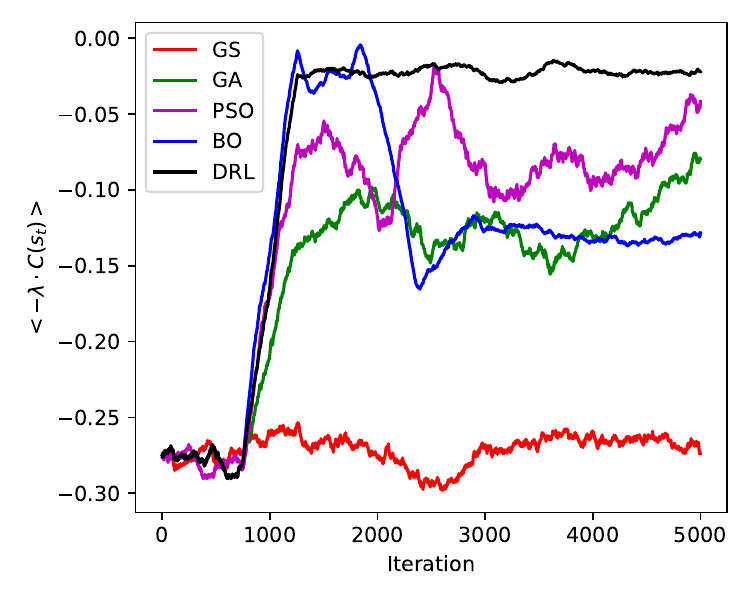}
         \caption{Penalty $-\lambda \cdot C(s_t)$}
         \label{fig:optimization-process-feasibility}
     \end{subfigure}
     \caption{Temporally averaged total reward (a) and penalty term (b) during the optimization process for different methods.}
    \label{fig:optimization-process-total}
\end{figure}

\begin{figure}[H]
\centerline{\includegraphics[width=\textwidth, scale = 1.0]{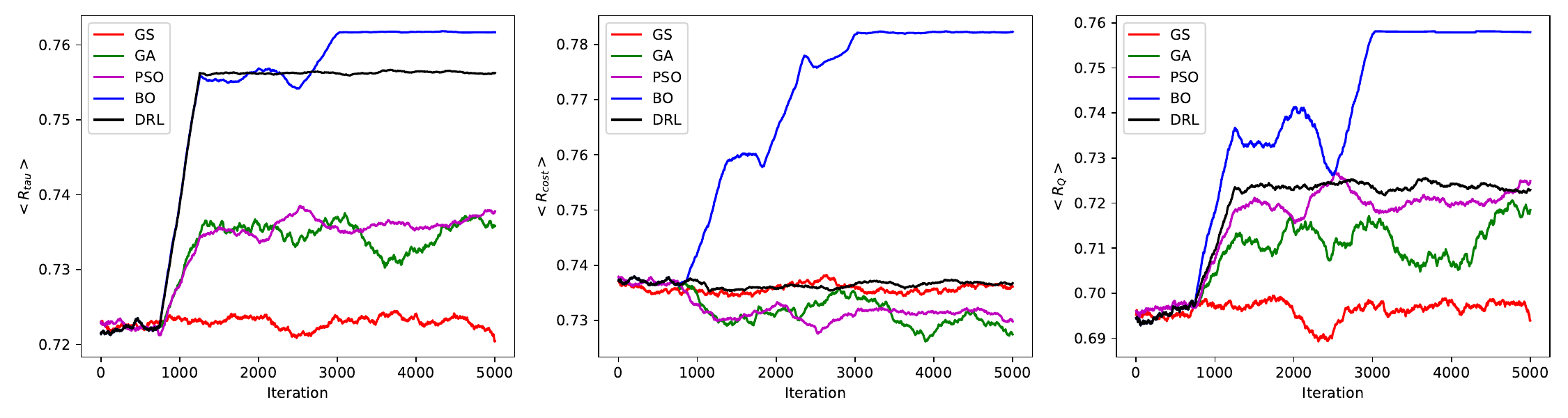}}
\caption{Temporal averaged reward corresponding to energy confinement time $\tau_\text{E}$, building cost $C$, and energy gain $Q$ during the optimization process.}
\label{fig:optimization-process-partial}
\end{figure}

In this section, we compare the optimization behavior of the baseline methods and the proposed DRL-based approach using the reward trajectories shown in Figures \ref{fig:optimization-process-total} and \ref{fig:optimization-process-partial}. Here, the candidate designs generated during the optimization process were collected, and their total reward $R_{\text{total}}$, penalty term $-\lambda \cdot C(s_t)$, and partial rewards for $\tau_E$, $C$, and $Q$ were evaluated. Figure \ref{fig:optimization-process-total} shows the temporal moving averages of the total reward and the penalty term over the iterations. As mentioned in the previous section, the offline stage corresponding to $n_{\text{iter}} < 1000$ is an offline data-collection phase used to initialize each optimization method. All methods exhibit similar rewards and penalties in the early iterations. Once the online stage begins, GA, PSO, BO, and the proposed DRL method all show clear improvements in total reward, whereas GS does not exhibit any improvement, as shown in Figure \ref{fig:optimization-process}. A similar tendency is observed in the penalty $-\lambda\cdot C(s_t)$ evaluation in Figure \ref{fig:optimization-process-feasibility}. In particular, the data-driven optimization methods progressively seek feasible designs, whereas GS continues to randomly explore the design space without incorporating constraints. Among all methods, the proposed DRL-based approach achieves the best optimization performance and finds more feasible solutions than the others. This indicates that the proposed method is not only effective in improving the design objectives but also more successful in exploring constraint-satisfying regions of the design space. These results demonstrate that the proposed method provides an effective framework for multi-objective tokamak reactor design optimization under operational constraints.

\begin{figure}[H]
\centerline{\includegraphics[width=0.75\textwidth, scale = 1.0]{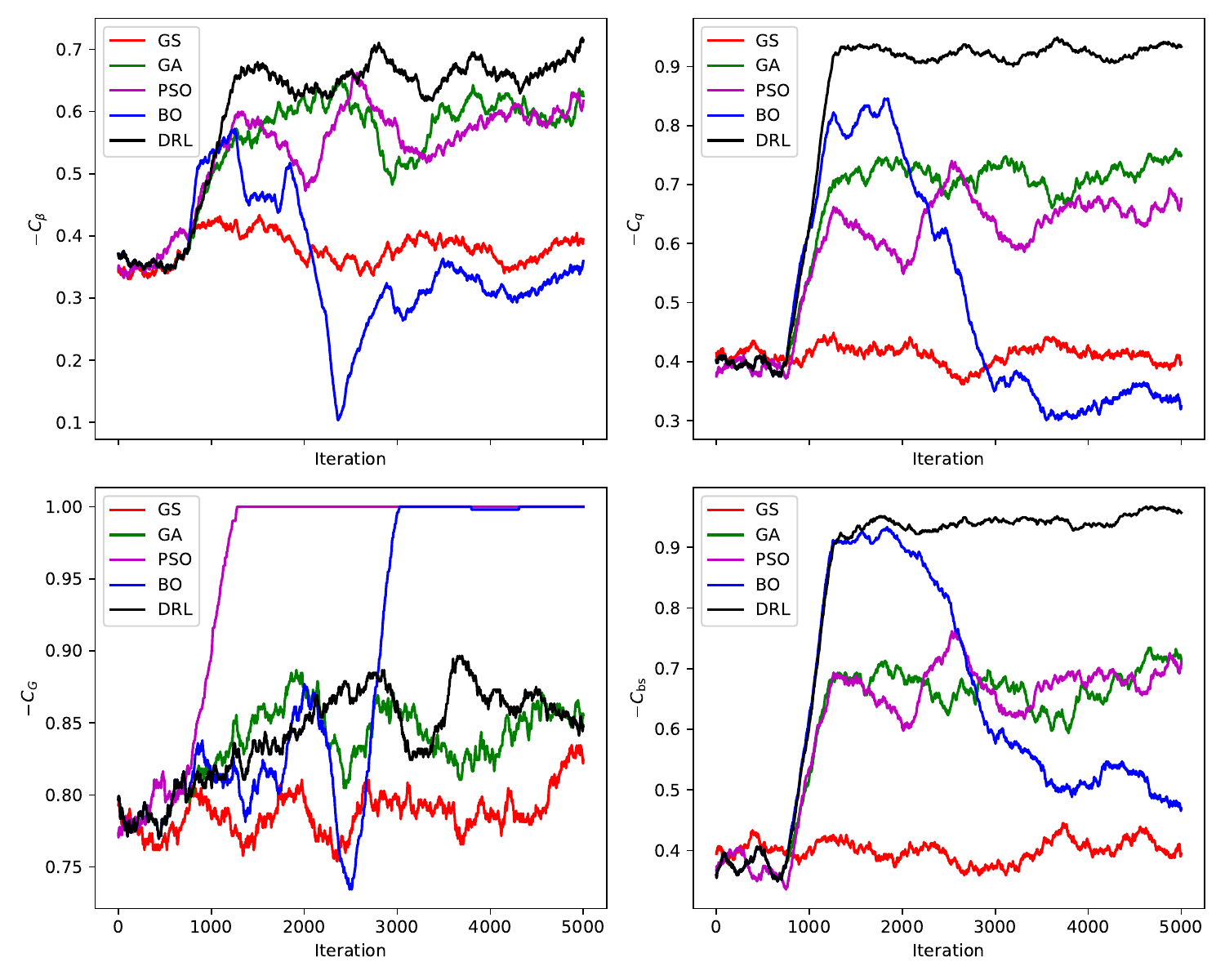}}
\caption{Temporal averaged penalty term corresponding to $\beta/\beta_{\text{Troyon}}$, $q_{\text{kink}}/q$, $\bar{n}/n_{\text{G}}$, and $f_{\text{op}}/f_{\text{bs}}$ during the optimization process.}
\label{fig:optimization-process-feasibility-partial}
\end{figure}

Figure \ref{fig:optimization-process-partial} decomposes the graph in \ref{fig:optimization-process-total} into partial rewards for the energy confinement time $\tau_\text{E}$, building cost $C$, and energy gain $Q$. Figure {\ref{fig:optimization-process-feasibility-partial}} describes the partial penalty term with respect to each operational constraint. In general, the rewards associated with $\tau_\text{E}$ and $Q$ improve over the optimization process for most data-driven methods. However, only BO shows a significant improvement in the building cost $C$. This result is explained by the fact that BO yields not only a greater cost reduction but also a significant decrease in the penalty term during optimization, as shown in \ref{fig:optimization-process-feasibility}. Figure {\ref{fig:optimization-process-feasibility-partial}} supports that BO has lower evaluation costs than others in the beta limit, and shows a significant decrease in the safety factor and bootstrap fraction limit over the iterations. This implies that part of the cost reduction achieved by BO is attributed to infeasible candidate designs. Moreover, this behavior reflects the trade-offs found in the multi-objective optimization problems. Since the reactor design is formulated as a constrained multi-objective optimization problem, an improvement in one objective does not necessarily mean balanced improvement across all objectives. Plus, the multiple constraints from the operational limits must be considered. From this perspective, the proposed approach yields the most balanced improvement across the competing objectives.

\begin{center}
\begin{tabular}{|P{3.0cm}|P{1.5cm}|P{1.5cm}|P{1.5cm}|P{1.5cm}|P{1.5cm}|}
 \hline
   Method & GS & GA & PSO & BO & \textbf{DRL} \\
 \hline
 Feasibility ratio & 0.62 \% & 3.00 \% & 1.78 \% & 2.30 \% & \textbf{11.00 \%} \\
 \hline
\end{tabular}
\captionof{table}{The proportions of feasible designs among all candidate designs explored during the optimization.}
\label{table:ratio-feasibility}
\end{center}

The proportions of the feasible solutions among all candidate designs explored are summarized in Table \ref{table:ratio-feasibility}. The proposed DRL-based method has the highest proportion of feasible designs among others, indicating that applying the RCPO enables the DRL approach to explore the design space in a more constraint-aware way than other approaches, without reducing exploration efficiency. Although BO shows highly cost-efficient candidate designs during the optimization process, its feasible-design ratio remains significantly lower than that of DRL.

\begin{figure}
\centerline{\includegraphics[width=0.75\textwidth, scale = 1.0]{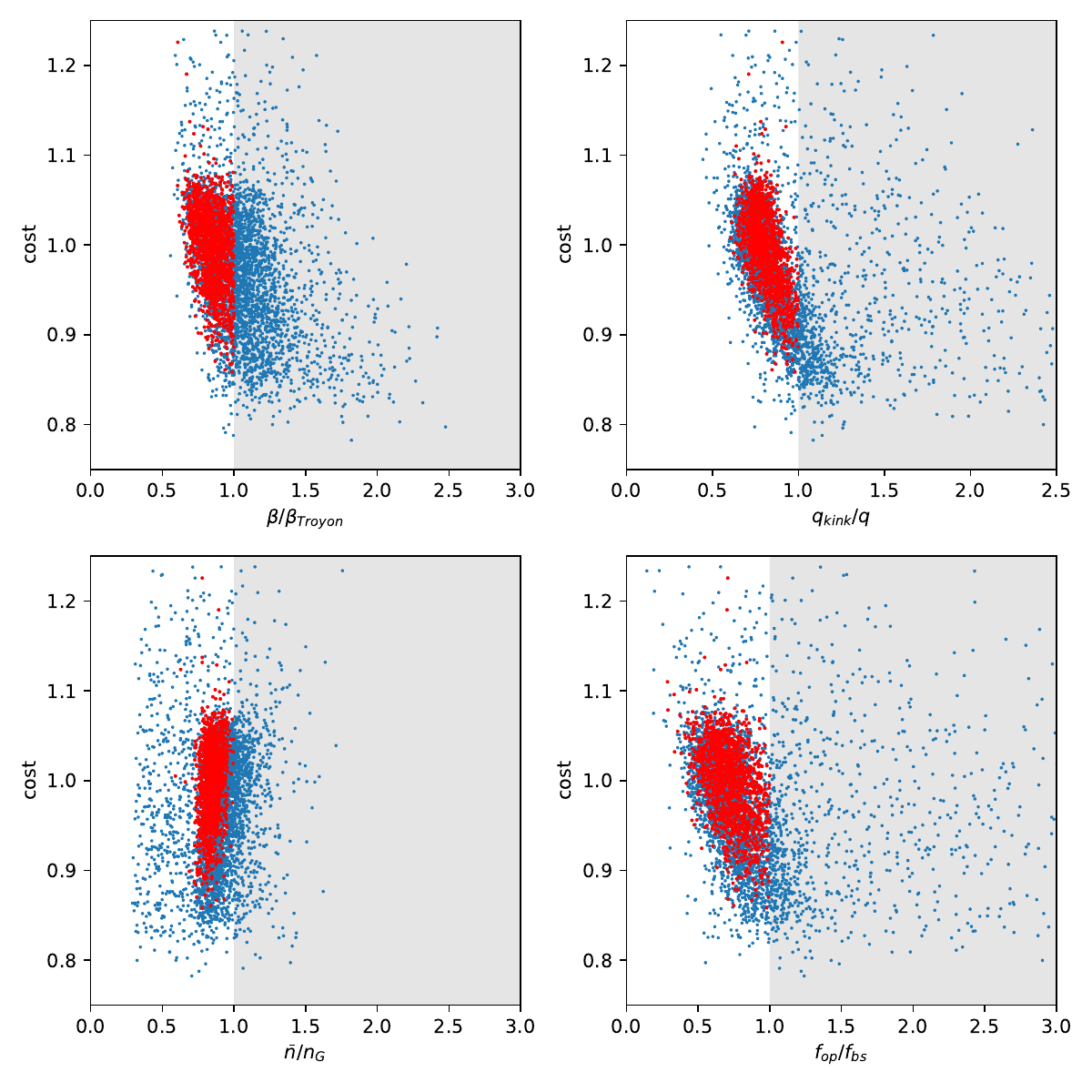}}
\caption{Distribution of the cost $C$ on $\beta/\beta_{\text{Troyon}}$, $q_{\text{kink}}/q$, $\bar{n}/n_{\text{G}}$, and $f_{\text{op}}/f_{\text{bs}}$.}
\label{fig:feasibility_cost}
\end{figure}

\begin{figure}
\centerline{\includegraphics[width=0.75\textwidth, scale = 1.0]{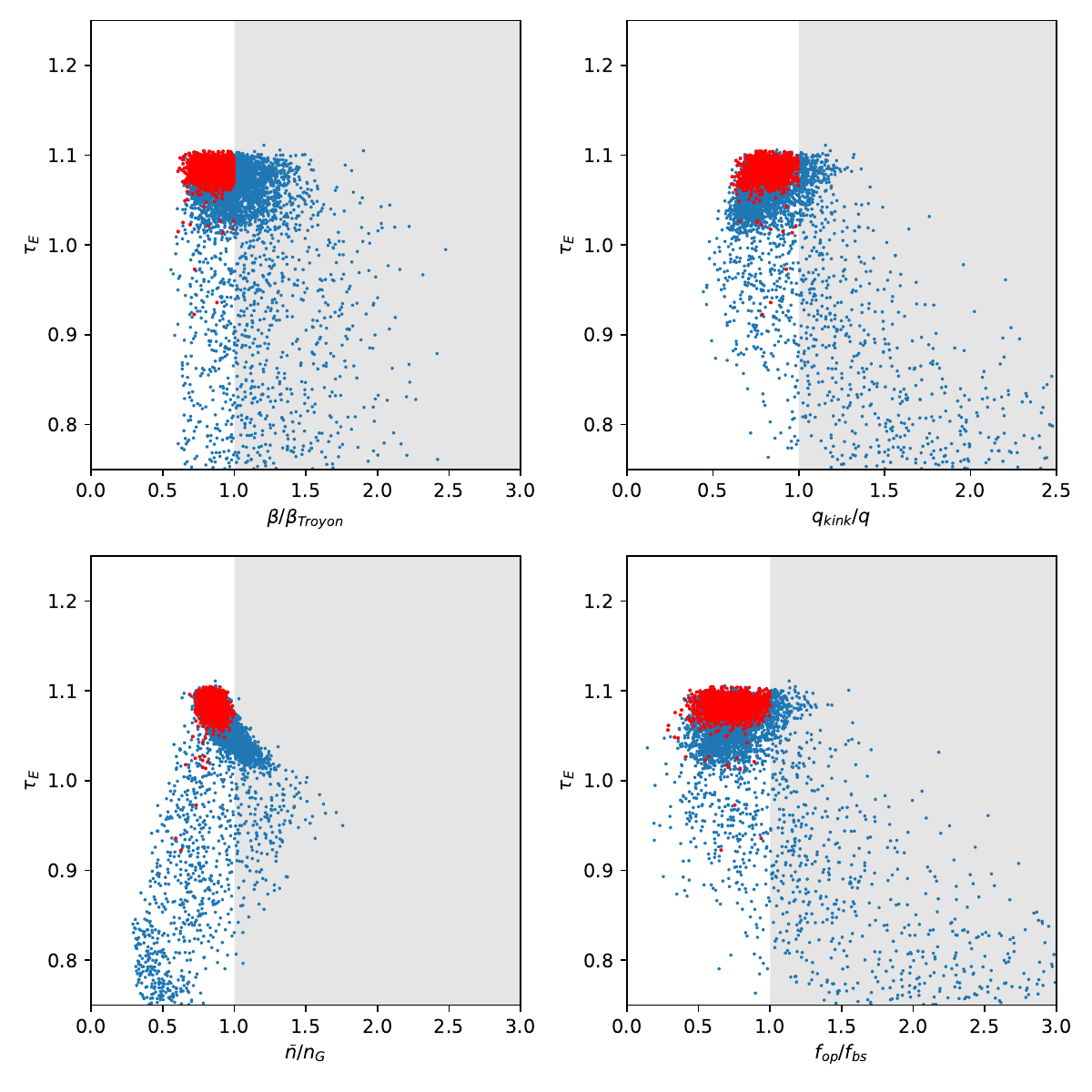}}
\caption{Distribution of the energy confinement time $\tau_\text{E}$ on $\beta/\beta_{\text{Troyon}}$, $q_{\text{kink}}/q$, $\bar{n}/n_{\text{G}}$, and $f_{\text{op}}/f_{\text{bs}}$.}
\label{fig:feasibility_tau}
\end{figure}

\begin{figure}
\centerline{\includegraphics[width=0.75\textwidth, scale = 1.0]{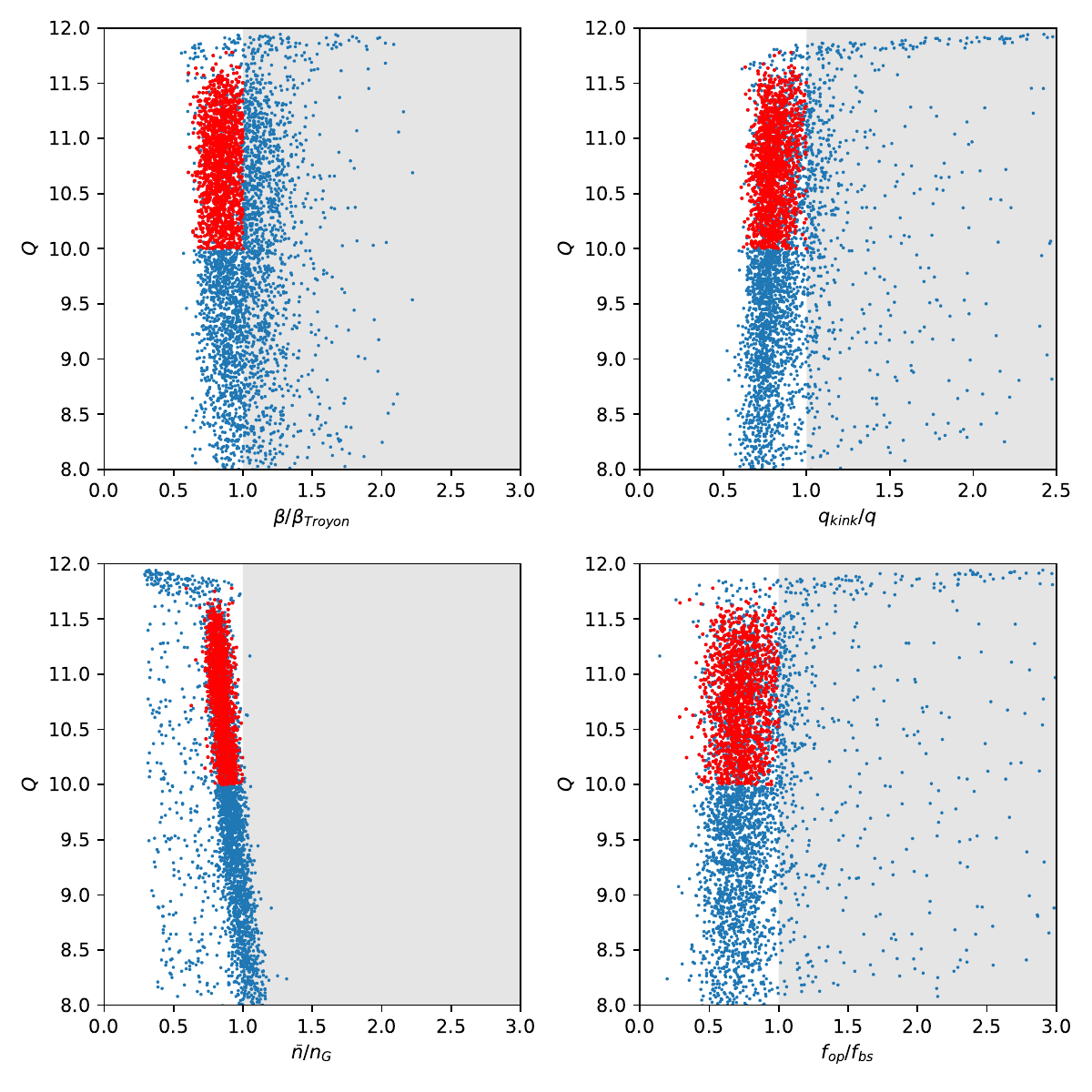}}
\caption{Distribution of the energy gain $Q$ on $\beta/\beta_{\text{Troyon}}$, $q_{\text{kink}}/q$, $\bar{n}/n_{\text{G}}$, and $f_{\text{op}}/f_{\text{bs}}$.}
\label{fig:feasibility_Q}
\end{figure}

To further examine the exploration efficiency of the proposed method, we plotted the distribution of the cost, energy confinement time, and energy gain on $\beta/\beta_{\text{Troyon}}$, $q_{\text{kink}}/q$, $\bar{n}/n_{\text{G}}$, and $f_{\text{op}}/f_{\text{bs}}$ from the collected designs from the DRL optimization, as in Figure \ref{fig:feasibility_cost}, \ref{fig:feasibility_tau}, and \ref{fig:feasibility_Q}. Here, the red dots represent feasible designs that satisfy all operational constraints, and the unshaded regions denote successful operational regimes that avoid the operational limits. We observed that the distribution in building cost follows a different pattern from that of the energy confinement time and energy gain. In particular, an excessive cost reduction tends to fall outside the unshaded region, indicating it may violate operational limits and supporting the previously mentioned discussion of cost reduction and feasibility. Plus, we can see that the proposed method explores a wide range of the design space while still identifying a comparatively large number of feasible design candidates. This supports that the method achieves both effective exploration and feasibility.

\subsection{Optimal design configurations}
Next, we compared the optimal design configuration that satisfies the operational constraints with the minimum building cost obtained by the different methods. Table \ref{table:comparison-objective} shows the comparison of the design objectives of the reference and the optimized designs. All methods successfully found optimal designs that reduce building cost while improving energy confinement time and energy gain. The DRL-optimized design is competitive with the best designs found by the other methods, demonstrating that the proposed framework can effectively identify high-performance fusion reactor designs under multiple design objectives and operational constraints.

\begin{center}
\begin{tabular}{|P{1cm}|P{1.25cm}|P{1.25cm}|P{1.25cm}|P{1.25cm}|P{1.25cm}|P{1.25cm}|}
 \hline
        & Ref \cite{freidberg2015designing} & GS & GA & PSO & BO & DRL \\
 \hline
    $C$ & 1.003 & 0.864 & 0.846 & 0.914 & 0.848 & 0.871 \\
    $\tau_\text{E}$ & 0.944 & 1.035 & 1.078 & 1.078 & 1.097 & 1.084 \\
    $Q$ & 10.380 & 10.394 & 10.958 & 11.017 & 11.828 & 11.000 \\
 \hline
\end{tabular}
\captionof{table}{The evaluation of the optimal design configurations with minimum building cost.}
\label{table:comparison-objective}
\end{center}

\begin{figure}
     \centering
     \begin{subfigure}[b]{0.32\textwidth}
         \centering
         \includegraphics[width=\textwidth]{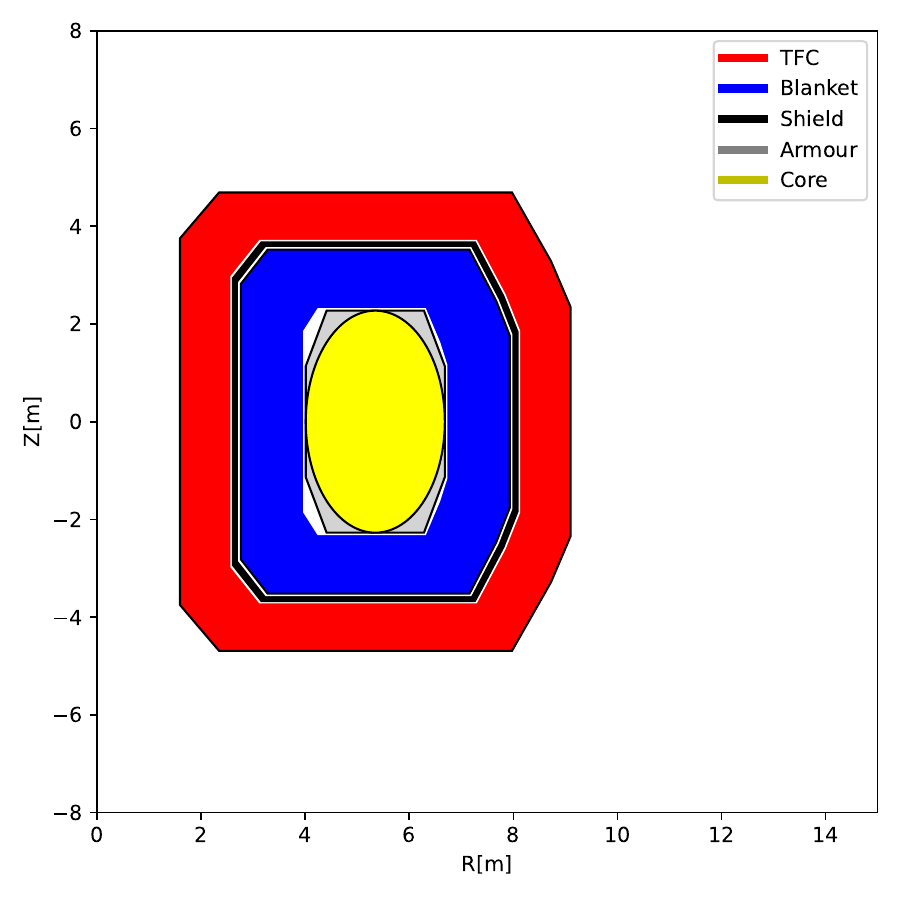}
         \caption{Reference \cite{freidberg2015designing}}
         \label{fig:poloidal-design-reference}
     \end{subfigure}
     \hfill
     \begin{subfigure}[b]{0.32\textwidth}
         \centering
         \includegraphics[width=\textwidth]{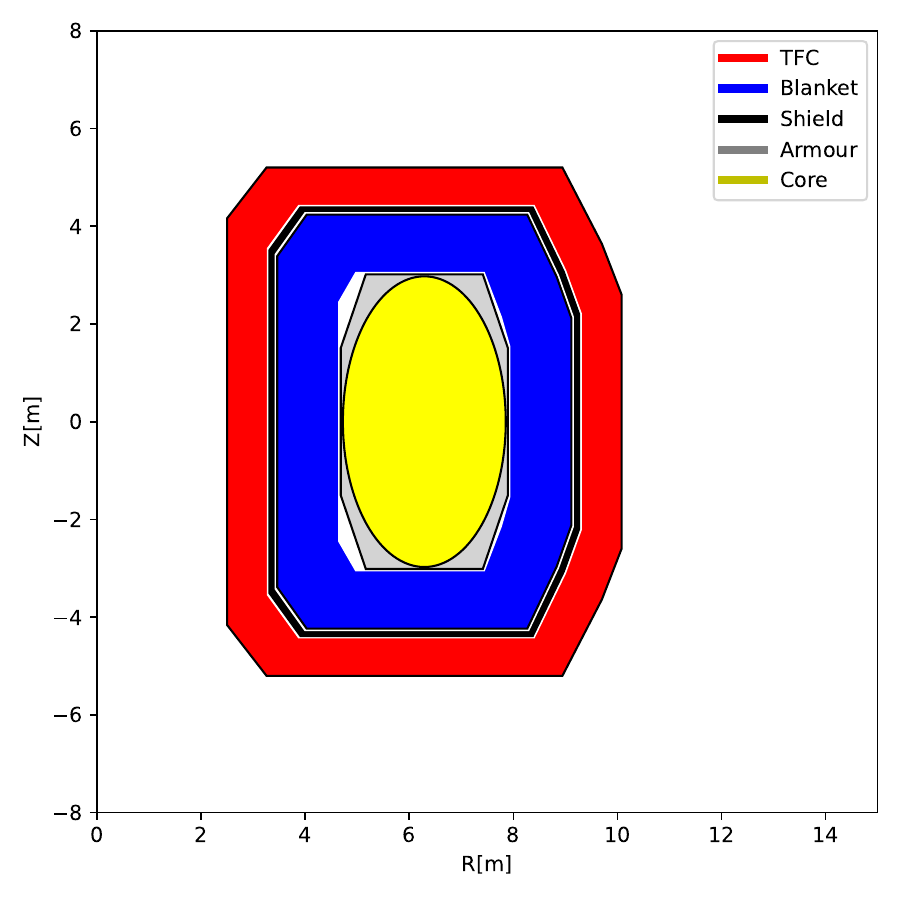}
         \caption{GS}
         \label{fig:poloidal-design-gridsearch}
     \end{subfigure}
     \hfill
     \begin{subfigure}[b]{0.32\textwidth}
         \centering
         \includegraphics[width=\textwidth]{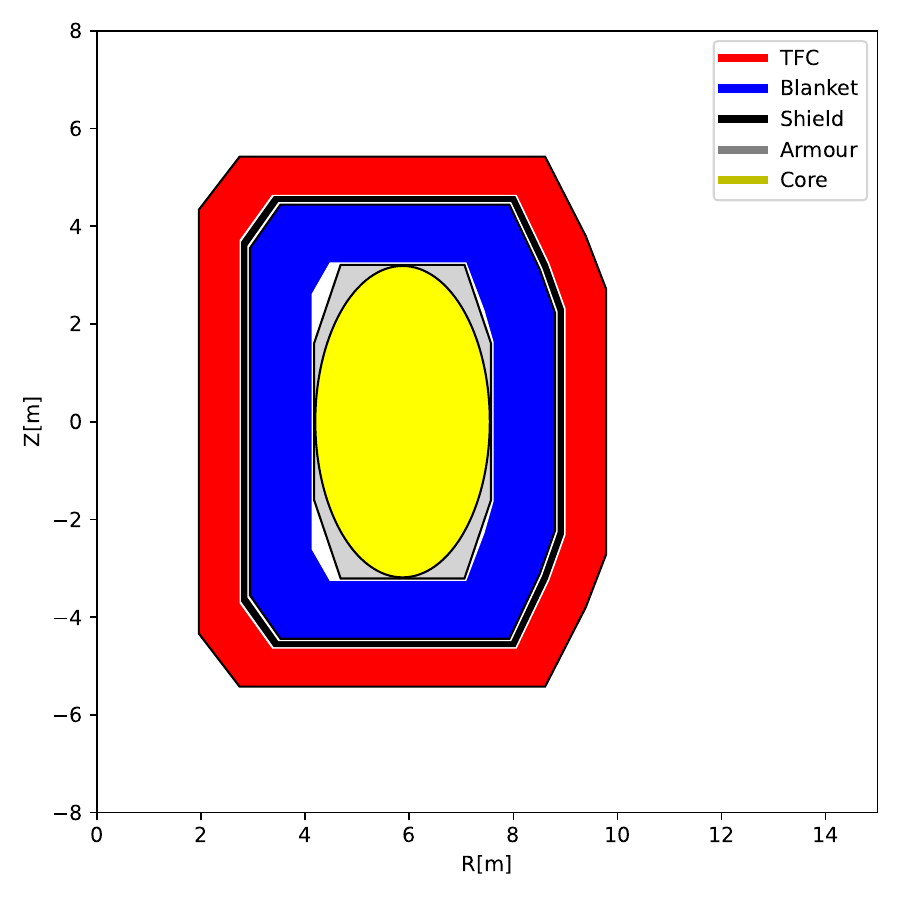}
         \caption{GA}
         \label{fig:poloidal-design-genetic}
     \end{subfigure}
     \vfill
     \begin{subfigure}[b]{0.32\textwidth}
         \centering
         \includegraphics[width=\textwidth]{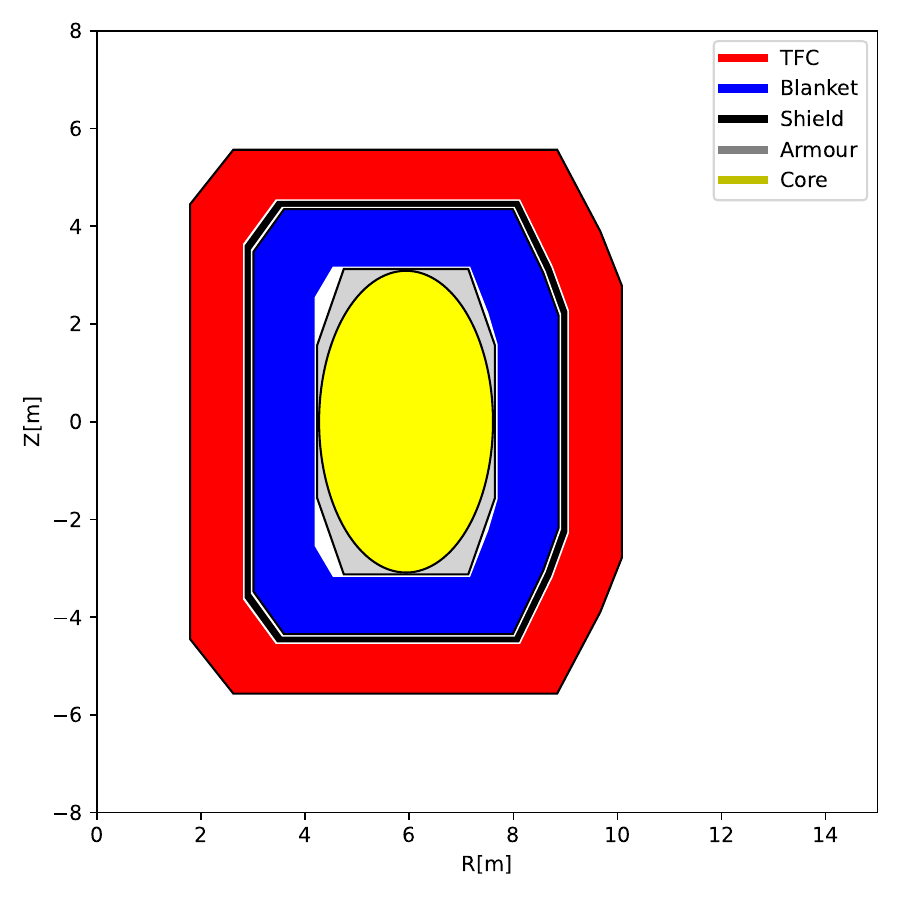}
         \caption{PSO}
         \label{fig:poloidal-design-particle}
     \end{subfigure}
     \hfill
     \begin{subfigure}[b]{0.32\textwidth}
         \centering
         \includegraphics[width=\textwidth]{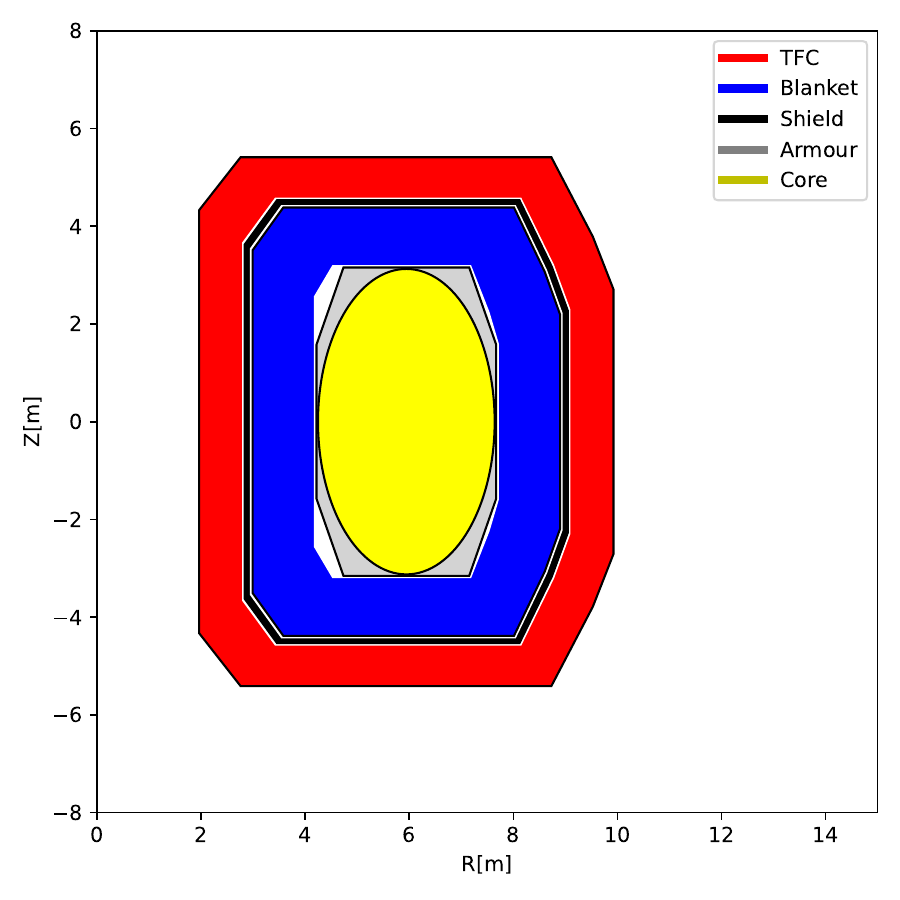}
         \caption{BO}
         \label{fig:poloidal-design-bayesian}
     \end{subfigure}
     \hfill
     \begin{subfigure}[b]{0.32\textwidth}
         \centering
         \includegraphics[width=\textwidth]{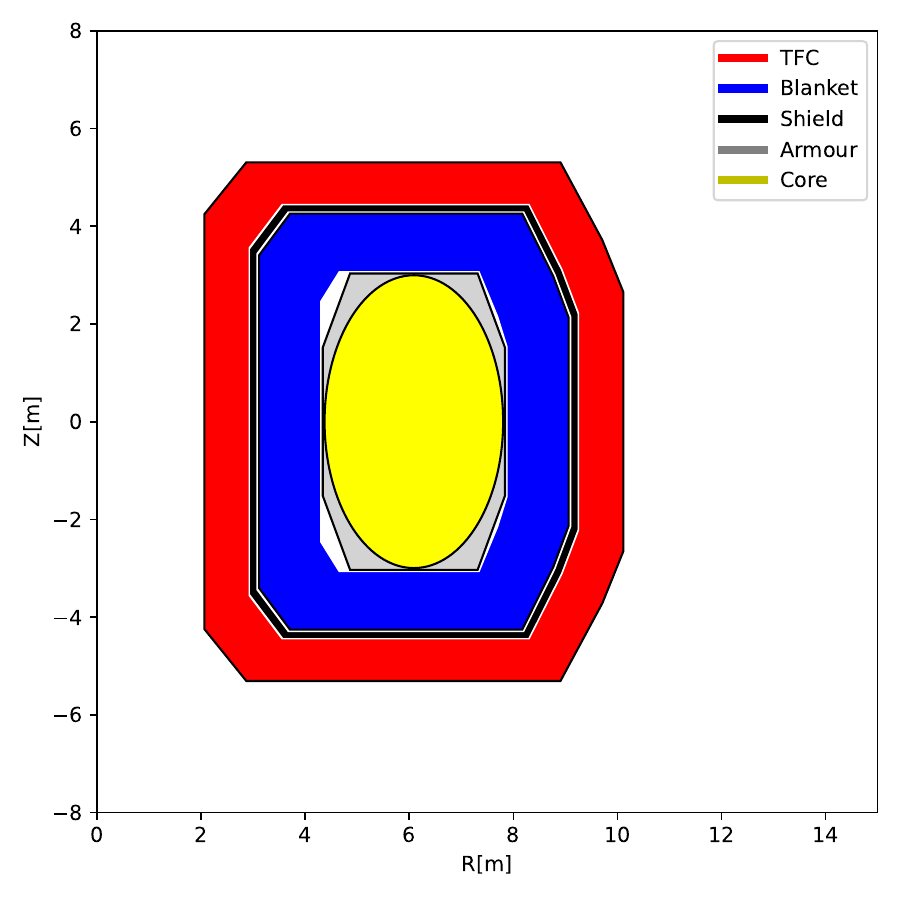}
         \caption{DRL}
         \label{fig:poloidal-design-rl}
     \end{subfigure}
        \caption{Comparison of the poloidal wall design configurations with different methods.}
        \label{fig:poloidal-design-comparison}
\end{figure}

Table \ref{table:design-parameter-comparsion} presents a detailed comparison of all design parameters for the reference and the optimized designs, and the corresponding poloidal cross-sectional designs are shown in Figure \ref{fig:poloidal-design-comparison}. One common feature of optimized designs is an increased size relative to the reference, along with higher fusion power. This result can be explained by the definition of the building cost $C$, which is given by the volume-to-generated-power ratio. In general, this ratio can be reduced either by decreasing both power and volume or by increasing both parameters. However, if both power and volume are reduced, this reduces the energy confinement time $\tau_E$ according to the scaling law \cite{yushmanov1990scalings}, leading to a decrease in the energy gain $Q$, which is inappropriate for our design objectives. Thus, it is reasonable to increase both power and size to reduce the cost while simultaneously improving energy confinement and energy gain. Besides, we can observe several common features across the optimal designs. First, the elongation is generally larger than that of the reference. Second, the optimized designs tend to operate at slightly higher temperatures and lower densities. This allows a smaller magnetic field while keeping $\beta$ within the achievable range and enabling a thinner coil by reducing the electromagnetic force on it. Lastly, the optimized designs generally exhibit a higher safety factor $q$ against the kink instability. Note that in this study, the kink instability constraint was set to the ideal limit $q_\text{kink}=2$ for comparison with the reference \cite{freidberg2015designing}, but this can be easily modified to a more practical limit ($\gtrsim3$) or combined with an external stability model.

\begin{center}
\begin{tabular}{|P{2.5cm}|P{1.5cm}|P{1.5cm}|P{1.5cm}|P{1.5cm}|P{1.5cm}|P{1.5cm}|}
 \hline
        & Reference  & GS & GA & PSO & BO & DRL \\
 \hline
 $R$(m) & 5.346 & 6.289 & 5.858 & 5.937 & 5.943 & 6.088 \\
 $a$(m) & 1.337 & 1.566 & 1.671 & 1.671 & 1.698 & 1.716\\
 $\Delta_{\text{blanket}}$(m) & 1.199 & 1.171 & 1.182 & 1.173 & 1.179 & 1.174\\
 $\Delta_{\text{shield}}$(m) & 0.1 & 0.1 & 0.1 & 0.1 & 0.1 & 0.1\\
 $\Delta_{\text{TF}}$(m) & 0.969 & 0.765 & 0.825 & 1.017 & 0.827 & 0.851\\
 $B_{\max}$(T) & 13.0 & 10.776 & 11.4 & 13.0 & 11.4 & 11.6\\
 $\kappa$ & 1.7 & 1.9 & 1.87 & 1.847 & 1.841 & 1.746\\
 $A$ & 4.0 & 4.017 & 3.505 & 3.552 & 3.5 & 3.547\\
 $H$ & 1.0 & 1.3 & 1.3 & 1.147 & 1.3 & 1.3 \\
 $\eta_{\text{thermal}}$ & 0.4 & 0.4 & 0.4 & 0.4 & 0.4 & 0.4\\
 $P_{E}$(MW) & 1000.0 & 1500.0 & 1472.8 & 1479.2 & 1500.0 & 1491.5 \\
 $P_{T}$(MW) & 2500.0 & 3750.5 & 3682.6 & 3698.4 & 3750.5 & 3729.1 \\
 $\beta$ & 4.125 & 4.742 & 4.983 & 3.701 & 4.979 & 4.617 \\
 $\tau_E$(s) & 0.944 & 1.035 & 1.078 & 1.078 & 1.100 & 1.084\\
 $I_p$(MA) & 14.296 & 12.170 & 13.327 & 14.080 & 13.182 & 14.599 \\
 $q$ & 1.544 & 2.247 & 2.363 & 2.515 & 2.383 & 2.043 \\
 $f_{\text{op}}$ & 0.841 & 0.588 & 0.751 & 0.738 & 0.533 & 0.672 \\
 $f_{\text{bs}}$ & 0.445 & 0.881 & 0.860 & 0.757 & 0.898 & 0.687 \\
 $Q_{\parallel}$(MW-T/m) & 500.0 & 567.2 & 577.0 & 663.84 & 582.0 & 582.9 \\
 $\bar{T}$(keV) & 14.0 & 14.0 & 14.4 & 14.5 & 15.0 & 14.4 \\
 $\bar{n}$($10^{20} m^{-3}$) & 1.43 & 1.30 & 1.24 & 1.23 & 1.20 & 1.23\\
 $\bar{p}$(atm) & 7.67 & 6.99 & 6.84 & 6.87 & 6.90 & 6.81 \\
 $Q$ & 10.38 & 10.39 & 10.96 & 11.02 & 11.83 & 11.00 \\
 Cost & 1.003 & 0.864 & 0.846 & 0.914 & 0.848 & 0.871 \\
 \hline
\end{tabular}
\captionof{table}{The comparison of design parameters from the reference \cite{freidberg2015designing} and the optimized designs.}
\label{table:design-parameter-comparsion}
\end{center}

\begin{center}
\begin{tabular}{|P{1.5cm}|P{1.25cm}|P{1.25cm}|P{1.25cm}|P{1.25cm}|P{1.25cm}|P{1.25cm}|}
 \hline
        & Ref \cite{freidberg2015designing} & GS & GA & PSO & BO & DRL \\
 \hline
    $\bar{n}/n_\text{G}$ & 0.560 & 0.823 & 0.815 & 0.770 & 0.819 & 0.779 \\
    $q_{\text{kink}}/q$ & 1.278 & 0.890 & 0.847 & 0.795 & 0.826 & 0.979 \\
    $\beta/\beta_{\text{Troyon}}$ & 0.942 & 0.928 & 0.955 & 0.751 & 0.935 & 0.987 \\
    $f_{\text{op}}/f_{\text{bs}}$ & 1.890 & 0.667 & 0.873 & 0.976 & 0.729 & 0.978 \\
 \hline
\end{tabular}
\captionof{table}{The comparison of the operational constraints from the reference and the optimized designs.}
\label{table:comparison-constraints}
\end{center}

\begin{figure}
\centerline{\includegraphics[width=0.75\textwidth, scale = 1.0]{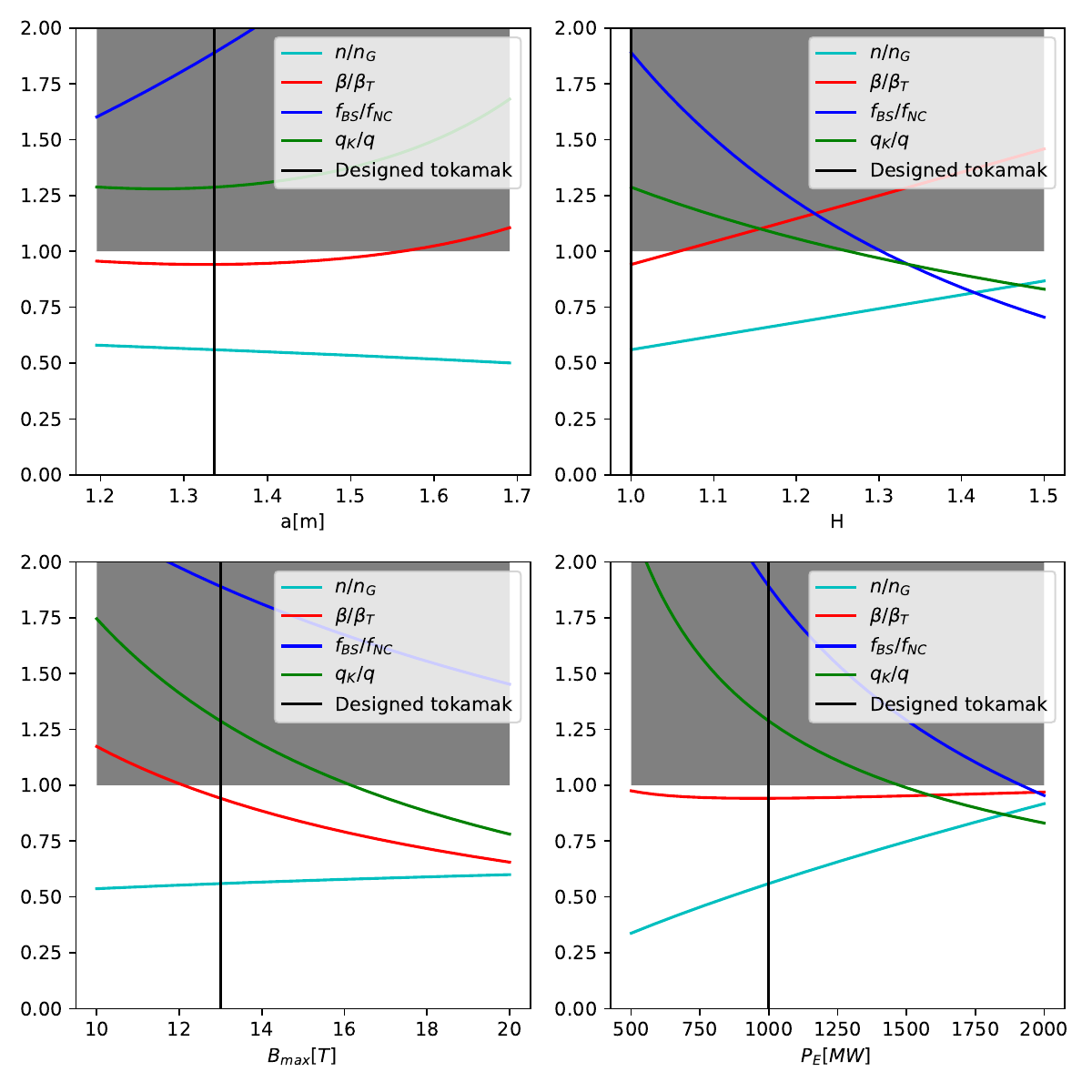}}
\caption{(Reference) Operational regime on the design parameters $a$,$H$,$B_{\max}$, and $P_\text{E}$, with the curves of $\frac{\bar{n}}{n_G}$,$\frac{\beta}{\beta_\text{Troyon}}$,$\frac{q_\text{kink}}{q}$, and $\frac{f_{\text{op}}}{f_{\text{bs}}}$. The unshaded region represents the successful operational regime.}
\label{fig:overall_performance_reference}
\end{figure}

\begin{figure}[H]
\centerline{\includegraphics[width=0.75\textwidth, scale = 1.0]{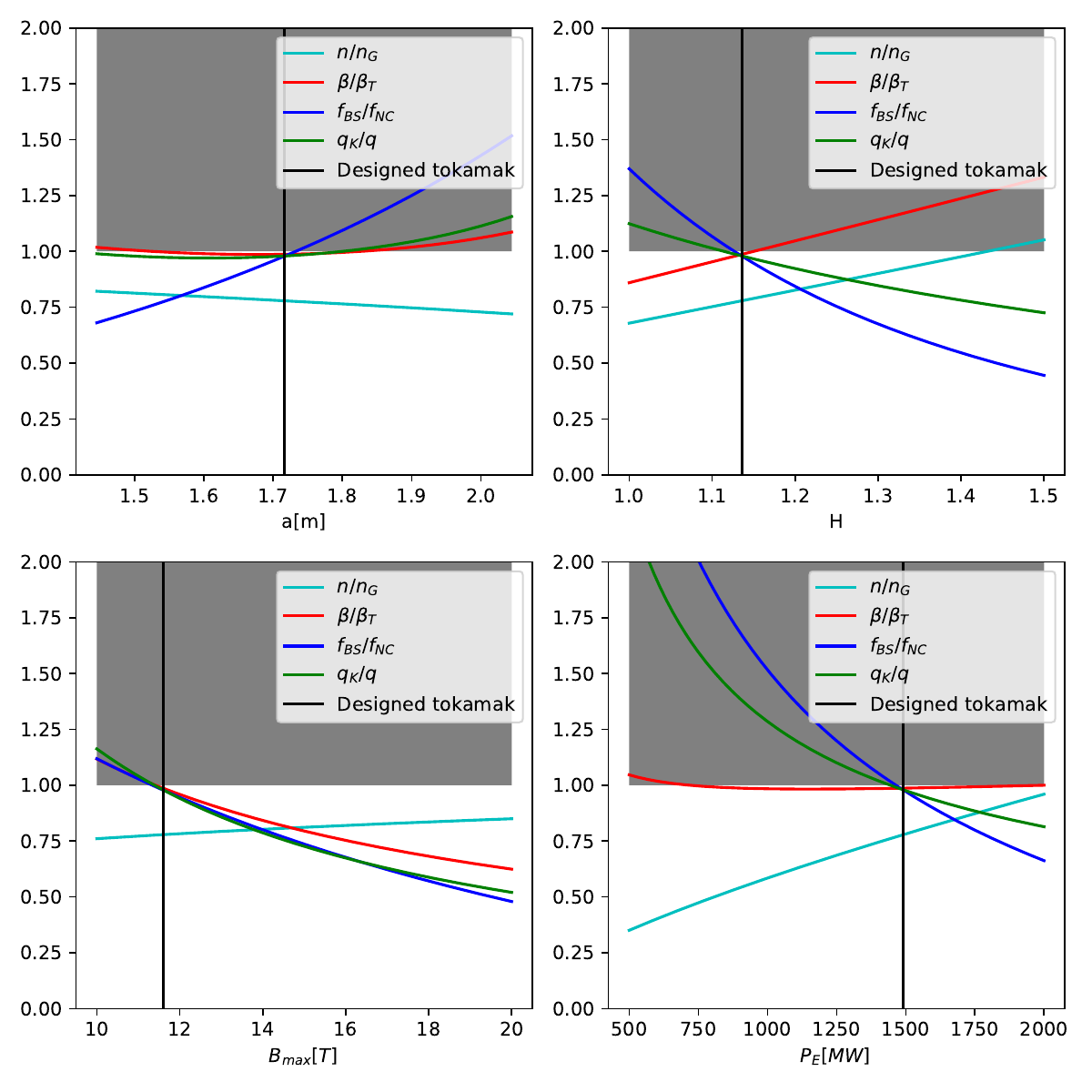}}
\caption{Operational regime on the design parameters $a$,$H$,$B_{\max}$, and $P_\text{E}$, with the curves of $\frac{\bar{n}}{n_\text{G}}$,$\frac{\beta}{\beta_\text{Troyon}}$,$\frac{q_\text{kink}}{q}$, and $\frac{f_{\text{op}}}{f_{\text{bs}}}$, optimized by DRL.}
\label{fig:overall_performance_rl}
\end{figure}

Finally, Table \ref{table:comparison-constraints} represents a comparison of the operational constraints between the reference and optimized designs. All fractions should be less than $1$ for steady-state operation. However, the reference case does not satisfy the safety factor condition for $m=2$ kink instability and the operational bootstrap condition. In contrast, the optimized designs satisfy these operational requirements. This is further illustrated in Figures \ref{fig:overall_performance_reference} and \ref{fig:overall_performance_rl}, which depict the operational regimes on the design parameters $a$,$H$,$B_{\max}$, and $P_\text{E}$, with the curves of $\frac{\bar{n}}{n_\text{G}}$,$\frac{\beta}{\beta_\text{Troyon}}$,$\frac{q_\text{kink}}{q}$, and $\frac{f_{\text{op}}}{f_{\text{bs}}}$ for the reference design and the optimal design by our proposed method. Here, the black vertical line in each plot indicates the status of the designed reactor. Each curve represents a change in stability with respect to input parameters and should be less than $1$. Thus, all curves should lie in the unshaded region. Figure \ref{fig:overall_performance_reference} is from \cite{freidberg2015designing}, which failed to find the safety regime for steady-state operation. However, the four plots in Figure {\ref{fig:overall_performance_rl}} from the proposed DRL-based approach lie in the unshaded region across all cases, indicating that this optimal design satisfied all constraints. 

Another observation from Table {\ref{table:comparison-constraints}} and Figure {\ref{fig:overall_performance_rl}} is that the DRL-optimized design lies closer to the boundary of the feasible region compared to other data-driven methods, except for the density limit. Operating close to these limits means that the available margins are used with small waste. For example, a higher value of $\beta$ is better as long as it does not exceed the Troyon limit since it yields a higher fusion power density. Similarly, $f_{\text{op}}/f_{\text{bs}} \approx 1$ means that the external current drive supplies only the fraction of the plasma current that the bootstrap current cannot, so no recirculating power is spent on current drive beyond what is needed. Note that in the case of density, a higher value is not necessarily better even if it is below the Greenwald limit, as shown by the fact that the Greenwald fraction of all optimized solutions is quite lower than 1, due to its dependency on current drive efficiency (See {\eqref{eq:plasma-current}} and {\eqref{eq:required-bs}}).

\section{Conclusion}
In this research, we investigated data-driven optimization methods for optimizing tokamak fusion reactor design under multiple operational constraints and competing design objectives. The goal was to minimize building cost and maximize the energy gain and energy confinement time, subject to the four major operational limits (density limit, beta limit, kink safety limit, and bootstrap fraction limit) required for stable operation. Motivated by the difficulty of efficiently searching for feasible, high-performing reactor configurations in a black-box design space, we proposed a deep reinforcement learning framework for fusion reactor design optimization that combines reward-constrained policy optimization (RCPO) with Proximal Policy Optimization (PPO). 

Our numerical experiments demonstrated that all data-driven optimization methods in this study were able to find feasible design configurations that satisfy all operational limit constraints while achieving lower cost parameters and higher confinement time and energy gain compared with the reference design. Among these approaches, however, our proposed method achieved the best optimization performance. First, this exhibited faster convergence and achieved higher average design evaluations during the optimization than the others. Since the reward includes both design objectives and constraint terms, this indicates that the proposed method effectively improves the design objectives while maintaining feasibility. Next, the RCPO-based DRL approach has the highest ratio of feasible solutions among all candidate designs explored during the optimization. This means that applying the RCPO enables the DRL approach to explore the design space in a more constraint-aware way than other approaches, without reducing exploration efficiency. Therefore, we conclude that the RCPO-based DRL approach provides a practical and effective way to optimize multi-objective, multi-constraint tokamak fusion reactor design, especially when feasibility and high performance are required simultaneously. 

Future work will focus on two main extensions. First, we will expand our design optimization framework to include additional design variables related to plasma profile shaping and wall material types. DRL can handle discontinuous variables and high-dimensional input data, thereby enabling optimization of 1D parameters and material composition. This extension would enable the simultaneous optimization of scalar design parameters, profile-related variables, and material-dependent properties, providing more practical and accurate insights for conceptualizing the optimal tokamak reactor design. This requires an integration of tools to model plasma responses based on higher-dimensional inputs, such as integrated plasma modeling \mbox{\cite{lee2021development}}, data-driven modeling for plasma confinement \mbox{\cite{nam2025machine}} and stability \mbox{\cite{kim2024disruption, kim2024enhancing}}, or physics-informed learning \mbox{\cite{seo2024solving, seo2024leveraging}}. Second, we will investigate multi-objective optimization from the perspective of Pareto optimality and user preference. In our work, multiple design objectives were combined through linear scalarization. This does not explicitly characterize the trade-off between the design objectives. Thus, the next step is to adopt a constrained multi-objective reinforcement learning approach \cite{huang2022constrained} that learns which design preference the policy network should optimize to satisfy the operational constraints. This will extend the proposed method beyond a single optimal reactor design toward a family of feasible Pareto-efficient design configurations, providing greater flexibility for building next-generation fusion reactors. 

\funding{This work was supported by a National Research Foundation of Korea (NRF) grant funded by the Korea government (MSIT) (Grant No. RS-2024-00346024).}

\roles{J.K. contributed to the design of this study, numerical experiments, analyses, and writing the manuscript. J.S. contributed to the supervision of this study, discussions, and writing the manuscript.}

\data{The codes that generate data in this paper are available at the GitHub repository, https://github.com/ZINZINBIN/RL-based-Reactor-Design-Optimization.}


\bibliographystyle{unsrt}
\IfFileExists{references.bib}{\bibliography{references}}{%
  \PackageWarningNoLine{manuscript}{references.bib not found; bibliography is omitted until uploaded}%
}

\end{document}